\UseRawInputEncoding
\documentclass{SciPost}

\hypersetup{
    colorlinks,
    linkcolor={red!50!black},
    citecolor={blue!50!black},
    urlcolor={blue!80!black}
}

\usepackage[bitstream-charter]{mathdesign}
\usepackage[utf8]{inputenc}
\usepackage[T1]{fontenc}
\usepackage{listingsutf8}
\usepackage{booktabs} 
\usepackage{siunitx} 
\usepackage{caption}  
\usepackage{tabularx}
\usepackage{ltablex}

\definecolor{codegreen}{rgb}{0,0.6,0}
\definecolor{codegray}{rgb}{0.5,0.5,0.5}
\definecolor{codepurple}{rgb}{0.84,0.15,}
\definecolor{backcolour}{rgb}{0.95,0.95, 0.95}
\definecolor{framecolor}{rgb}{0.8,0.8, 0.8}

\lstdefinestyle{mystyle}{
    backgroundcolor=\color{backcolour},   
    commentstyle=\color{codegreen},
    keywordstyle=\color{magenta},
    numberstyle=\tiny\color{codegray},
    stringstyle=\color{codepurple},
    basicstyle=\ttfamily\footnotesize,
    breakatwhitespace=false,         
    breaklines=true,                 
    captionpos=t,                    
    keepspaces=true,                 
    numbers=none,                    
    numbersep=5pt,                  
    showspaces=false,                
    showstringspaces=false,
    showtabs=false,                  
    tabsize=2
}

\lstdefinelanguage{Toml}{
    comment = [l]{\#},
    keywords = {true, false},
    morestring = [b]{"}
}

\lstdefinelanguage{Julia}%
  {morekeywords={abstract,break,case,catch,const,continue,do,else,elseif,%
      end,export,false,for,function,immutable,import,importall,if,in,%
      macro,module,otherwise,quote,return,switch,true,try,type,typealias,%
      using,while},%
   sensitive=true,%
   alsoother={},%
   morecomment=[l]\#,%
   morecomment=[n]{\#=}{=\#},%
   morestring=[s]{"}{"},%
   morestring=[m]{'}{'},%
}[keywords,comments,strings]%

\DeclareSymbolFont{usualmathcal}{OMS}{cmsy}{m}{n}
\DeclareSymbolFontAlphabet{\mathcal}{usualmathcal}

\fancypagestyle{SPstyle}{
\fancyhf{}
\lhead{\colorbox{scipostblue}{\bf \color{white} ~SciPost Physics Codebases }}
\rhead{{\bf \color{scipostdeepblue} ~Submission }}

\fancyfoot[C]{\textbf{\thepage}}
}

\begin{document}

\pagestyle{SPstyle}

\begin{center}{\Large \textbf{\color{scipostdeepblue}{
XDiag: Exact Diagonalization for Quantum Many-Body Systems\\
}}}\end{center}

\begin{center}\textbf{
Alexander Wietek\textsuperscript{1$\star$}, 
Luke Staszewski\textsuperscript{1},
Martin Ulaga\textsuperscript{1},
Paul L. Ebert\textsuperscript{1},
Hannes Karlsson\textsuperscript{1},
Siddhartha Sarkar\textsuperscript{1},
Leyna Shackleton\textsuperscript{2},
Aritra Sinha\textsuperscript{1},
Rafael D. Soares\textsuperscript{1}
}\end{center}

\begin{center}
{\bf 1} Max Planck Institute for the Physics of Complex Systems, Nöthnitzer Straße 38, Dresden 01187, Germany

{\bf 2} Center for Computational Quantum Physics, Flatiron Institute, 162 Fifth Avenue, New York, New York 10010, USA

{\bf 3} Department of Physics, Massachusetts Institute of Technology, Cambridge, Massachusetts 02139, USA
\\

$\star$ \href{mailto:awietek@pks.mpg.de}{\small awietek@pks.mpg.de}\,,\quad
\end{center}

\section*{\color{scipostdeepblue}{Abstract}}
\textbf{\boldmath{%
Exact diagonalization (ED) is a cornerstone technique in quantum many-body physics, enabling precise solutions to the many-body Schrödinger equation for interacting quantum systems~\cite{Oitmaa1978,Dagotto1994,Weisse2008,Prelovsek1994}. Despite its utility in studying ground states, excited states, and dynamical behaviors, the exponential growth of the Hilbert space with system size presents significant computational challenges. We introduce XDiag, an open-source software package designed to combine advanced and efficient algorithms for ED with and without symmetry-adapted bases with user-friendly interfaces. Implemented in C++ for computational efficiency and wrapped in Julia for ease of use, XDiag provides a comprehensive toolkit for ED calculations. Key features of XDiag include the first publicly accessible implementation of sublattice coding algorithms for large-scale spin system diagonalizations, efficient Lin table algorithms for symmetry lookups, and random-hashing techniques for distributed memory parallelization. The library supports various Hilbert space types (e.g., spin-1/2, electron, and t-J models), facilitates symmetry-adapted block calculations, and automates symmetry considerations. The package is complemented by extensive documentation, a user guide, reproducible benchmarks demonstrating near-linear scaling on thousands of CPU cores, and over 20 examples covering ground-state calculations, spectral functions, time evolution, and thermal states. By integrating high-performance computing with accessible scripting capabilities, XDiag allows researchers to perform state-of-the-art ED simulations and explore quantum many-body phenomena with unprecedented flexibility and efficiency.
}}

\vspace{\baselineskip}

\noindent\textcolor{white!90!black}{%
\fbox{\parbox{0.975\linewidth}{%
\textcolor{white!40!black}{\begin{tabular}{lr}%
  \begin{minipage}{0.6\textwidth}%
    {\small Copyright attribution to authors. \newline
    This work is a submission to SciPost Physics Codebases. \newline
    License information to appear upon publication. \newline
    Publication information to appear upon publication.}
  \end{minipage} & \begin{minipage}{0.4\textwidth}
    {\small Received Date \newline Accepted Date \newline Published Date}%
  \end{minipage}
\end{tabular}}
}}
}


\vspace{10pt}
\noindent\rule{\textwidth}{1pt}
\tableofcontents
\noindent\rule{\textwidth}{1pt}
\vspace{10pt}


\section{Introduction}
\label{sec:intro}
Exact diagonalization (ED) is a fundamental technique in quantum many-body physics~\cite{Oitmaa1978,Dagotto1994,Weisse2008,Prelovsek1994,Lin1990,Schnack2023,Schulz1996,Hasegawa1989,Leung1993,Vary2009,Maris2010,Laeuchli2019}. It provides precise solutions to the Schr\"{o}dinger equation for interacting quantum systems. This method is crucial for understanding the ground state properties, excited states, and dynamical behaviors of quantum systems and serves as a benchmark for other approximate methods. However, the exponential growth of the Hilbert space with system size poses significant computational challenges, necessitating efficient algorithms and high-performance computing resources.

As a fundamental technique for studying quantum matter, open-source community codes are important to advance our respective fields. Indeed, several ED software packages have so far been released, each with its respective strengths. Early on, the ALPS libraries~\cite{Albuquerque2007} provided the community with a reliable and well-tested package. Another early versatile ED package in frequent use is TITPACK~\cite{titpack}. QuSpin~\cite{Weinberg2017,Weinberg2019} is an ED library that excels in its versatility and user-friendliness by offering a Python interface and is also widely adopted by the community. Spinpack~\cite{spinpack} is a versatile library for spin systems that offers well-scaling distributed memory parallelization. Similarly, HPhi~\cite{Kawamure2017} has been used to perform Exact Diagonalization on large distributed-memory clusters and provides excellent tools for finite-temperature simulations. Highly scalable distributed memory parallelizations for spin systems have been presented recently \cite{Westerhout2023,Schaefer2025a}. Also, EDIPack was released featuring efficient ED algorithms for quantum impurity models~\cite{Amaricci2022}. Versatile and powerful libraries for ED of interacting electrons are further the Pomerol library~\cite{Pomerol2015} and the EDLib library~\cite{Isakov2018} together and the libcommute package~\cite{Krivenko2022}, which feature generic fermionic interactions. 

A particular strength of ED is to perform calculations in a symmetry-adapted basis of the Hilbert space. Not only does employing symmetries reduce the resource requirements and speed up calculations, but it also allows extracting physical insights directly by investigating the spectral structure, also known as tower-of-states (TOS) analysis~\cite{Wietek2017}. Efficient implementations of symmetry-adapted bases in ED, however, require specialized algorithms to be performed optimally. The so-called Lin tables~\cite{Lin1990} have been proposed early as an efficient algorithm for indexing Hilbert spaces of spin systems with spin conservation and translational symmetry. Recently, these indexing ideas have been significantly improved to arbitrary spin-$S$ by the library DanceQ~\cite{Schaefer2025a,Schaefer2025b}, which also offers highly efficient distributed memory parallelization. Another route for (memory-)efficient use of space group symmetries is the so-called sublattice coding techniques~\cite{Schulz1996,Weisse2013,Wietek2018}, which have been shown to allow for diagonalizations of spin $S=1/2$ of up to $N=50$ particles~\cite{Wietek2018,Wu2024}. Moreover, several works have also successfully constructed full lattice and spin rotational symmetric bases in ED simulations~\cite{Schnalle2010,Heitmann2019}

While these techniques offer significant computational advantages, only a few existing open-source software packages offer access to these algorithms. With XDiag, we aim to combine both the efficiency of these advanced algorithms for working in symmetry-adapted bases with the user-friendliness offered by higher-level programming languages. While our core library is implemented in C++ for efficiency and expressiveness, we use Julia as a high-level scripting language and provide a comprehensive wrapper. This also has the advantage that, in principle, only one language needs to be employed to run both simulations and data analysis, including visualizations. Our API is strongly inspired by the ITensor library~\cite{itensor,itensor-r0.3} for tensor network simulations (ITensors.jl, ITensorMPS.jl), such that the overall high-level user experience is rather similar. Among numerous other smaller algorithms, XDiag features the first publicly accessible implementation of sublattice coding algorithms~\cite{Wietek2018} for large-scale diagonalization of spin systems, implementations of various forms of Lin tables~\cite{Lin1990} for fast symmetry lookup, and several random-hashing algorithms for workload balancing in distributed memory parallelization~\cite{Wietek2018}. Moreover, we provide algorithms to analyze quantum numbers like momentum and particle numbers of many-body operators to automatize symmetry considerations in the many-body context. Input and output are handled conveniently using TOML~\cite{marzerToml} and HDF5~\cite{hdf5} files. In the past, XDiag has already been employed in several publications~\cite{Wietek2020,Wietek2021,Shackleton2021,Feng2022,ZWang2024,Ulaga2024,Wietek2024,Sanyal2024,Staszewski2024,Haldar2024}. Details on the algorithms implemented can be found in section~\ref{sec:implementation}. 

To make the XDiag library accessible to a general audience, we provide the following resources.

\paragraph{Documentation:} We define the API for both the Julia and C++ versions in our online documentation~\cite{documentation}, which features detailed instructions for all functions and classes defined, including usage examples for individual function calls. The API also defines the set of exported symbols if the C++ library is compiled as a shared object.

\paragraph{User guide:} Alongside the documentation, we provide a user guide introducing the concepts and logic used in XDiag in an accessible way. This includes installation and compilation instructions with a walk-through guide to using XDiag. The guide is accessible via the online documentation~\cite{documentation} and presented in section~\ref{sec:usage_guide}. 

\paragraph{Examples:} An extensive set of $20+$ examples, which showcase the versatility of XDiag is provided. Examples range from ground state calculations of spin models to dynamical response functions of electronic systems, entanglement properties compared to conformal field theories, level statistics of (non-)integrable models, time-evolution and thermodynamic properties of various systems. The examples together with visualization scripts are available in the online documentation~\cite{documentation} and presented in section~\ref{sec:examples}. 

\paragraph{Benchmarks:} We benchmarked our implementation of various methods and provided the code to run these benchmarks, such that they can be easily reproduced. This includes scaling analysis for both the shared-memory as well as distributed-memory parallelization, which is shown to exhibit almost linear behavior in all test cases and system sizes up to several thousands of CPU cores. We describe the details of these benchmarks in section~\ref{sec:benchmarks}.

\paragraph{Unit tests:} While our unit tests are mainly implemented to assure the correctness of XDiag, they can also serve as a potential source of information on how to use the code. The unit tests are run as continuous integration testing on GitHub for Linux and MacOS architectures, both for the shared- and distributed-memory case. 

\section{Installation}
\label{sec:installation}

\subsection{Julia library}
XDiag can be conveniently installed via the Julia package manager. First, a Julia interpreter needs to be opened from a command line using \lstinline{julia}. Then, the "package mode" can be entered by typing \lstinline{]}, and XDiag is installed using \lstinline{add XDiag}. In summary:
\begin{lstlisting}[language=Julia]
$ julia
julia> ]
pkg> add XDiag
\end{lstlisting}

\subsection{C++ library}
The first step in employing the C++ version is to compile the library. The source code can be obtained from GitHub~\cite {github} by cloning using \lstinline{git}~\cite{gitscm}.
\begin{lstlisting}[language=bash]
cd /path/to/where/xdiag/should/be
git clone https://github.com/awietek/xdiag.git
\end{lstlisting}
The compilation and installation are then performed with \lstinline{CMake}~\cite{cmake}. 
\begin{lstlisting}[language=bash]
cd xdiag
cmake -S . -B build
cmake --build build
cmake --install build
\end{lstlisting}
By default, the resulting library is installed at \lstinline{/path/to/where/xdiag/should/be/install}. There are various options when compiling, including optimizations that can be used. For more details on the compilation process, we refer to the compilation guide in the documentation~\cite{documentation}.

\subsection{Writing application code}
To employ the XDiag library, an application code is written. The simplest application using XDiag is a \lstinline{hello_world} program.

\noindent
\begin{minipage}[t]{.48\textwidth}
\begin{lstlisting}[language=C++,title=C++,belowcaptionskip=-10pt]
#include <xdiag/all.hpp>
using namespace xdiag;
int main() try {
    say_hello();
} catch (Error e) {
    error_trace(e);
}
\end{lstlisting}
\end{minipage}
\hfill
\begin{minipage}[t]{.48\textwidth}
\begin{lstlisting}[language=Julia,title=Julia,belowcaptionskip=-10pt]
using XDiag
say_hello()
\end{lstlisting}
\end{minipage}

\noindent The function \lstinline{say_hello()} prints out a welcome message, which also contains information on which exact XDiag version is used. We would like to emphasize the \lstinline{try / catch} block in the C++ version. XDiag implements a traceback mechanism for runtime errors, which is activated by the \lstinline{error_trace} function. While not strictly necessary here, it is a good practice to employ this.

\subsection{Application compilation}\label{sec:applicationCompilation}
In C++, now that the application program is written, we need to set up the compilation instructions using CMake. To do so, we create a second file called \lstinline{CMakeLists.txt} in the same directory.

\begin{lstlisting}[language=bash]
cmake_minimum_required(VERSION 3.19)
project(hello_world)
find_package(xdiag REQUIRED HINTS "/path/to/where/xdiag/should/be/install")
add_executable(main main.cpp)
target_link_libraries(main PRIVATE xdiag::xdiag)  
\end{lstlisting}
You should replace \lstinline{/path/to/where/xdiag/should/be/install} with the appropriate directory where your XDiag library is installed after compilation. We then compile the application code,
\begin{lstlisting}
cmake -S . -B build
cmake --build build
\end{lstlisting}
and finally run our first XDiag application.
\begin{lstlisting}
./build/main
\end{lstlisting}

\section{Usage guide}
\label{sec:usage_guide}

After installation or library compilation, the user is ready to run the first calculation using XDiag. For the purpose of this usage guide, our immediate goal is to determine the ground state energy of a $S=1 / 2$ Heisenberg model on a 1D chain lattice with periodic boundary conditions,
\begin{equation}
H=J \sum_{\langle i, j\rangle} \mathbf{S}_{i} \cdot \mathbf{S}_{j},
\label{eq:heisenberg_model}
\end{equation}
where $\mathbf{S}_{i}=\left(S_{i}^{x}, S_{i}^{y}, S_{i}^{z}\right)$ denotes the vector of spin matrices at a given site $i$. The notation $\langle i, j\rangle$ refers to summation over neighboring sites $i$ and $j$. In the following, we will discuss how to set up such a calculation and give an overview of further functionality in XDiag.

\subsection{Hilbert spaces}
\label{subsec:hilbert_space}
\noindent The first object to define before any computation is the Hilbert space on which our model will be defined. For this, we create an object of type \lstinline{Spinhalf} and hand as an argument the number of physical sites \lstinline{N}.

\noindent
\begin{minipage}[t]{.48\textwidth}
\begin{lstlisting}[language=C++,title=C++,belowcaptionskip=-10pt]
int N = 8;
auto hs = Spinhalf(N);
\end{lstlisting}
\end{minipage}
\hfill
\begin{minipage}[t]{.48\textwidth}
\begin{lstlisting}[language=Julia,title=Julia,belowcaptionskip=-10pt]
N = 8
hs = Spinhalf(N)
\end{lstlisting}
\end{minipage}
We would like to know which spin configurations the Hilbert space is made up of. To do so, we can iterate over the Hilbert space and print out the configurations and the total Hilbert space dimension.

\noindent
\begin{minipage}[t]{.48\textwidth}
\begin{lstlisting}[language=C++,title=C++,belowcaptionskip=-10pt]
for (auto spins: hs) {
    Log("{}", to_string(spins));
    Log("{}", index(hs, spins));
}
Log("dim: {}", size(hs));
\end{lstlisting}
\end{minipage}
\hfill
\begin{minipage}[t]{.48\textwidth}
\begin{lstlisting}[language=Julia,title=Julia,belowcaptionskip=-10pt]
for spins in hs
    println(to_string(spins))
    println(index(hs, spins))
end
println("dim: ", size(hs))
\end{lstlisting}
\end{minipage}
XDiag features a convenient way to write logs in C++, with the \lstinline{Log} class. The first argument to \lstinline{Log()} is a format string. In C++, we use the \lstinline{fmt} library~\cite{fmtlib} to be able to write formatted output. The second argument turns our spins into a string. \lstinline{spins} is of type \lstinline{ProductState}, whose configuration on each site can be individually addressed by using the \lstinline{[]} operator. The opposite task of computing the index of a given \lstinline{ProductState} within the block basis can be addressed using the \lstinline{index} function. An important difference between C++ and Julia is that indices are counted starting from \lstinline{0} in C++ and \lstinline{1} in Julia. Hence, also in the above code snippet, the C++ version will start counting the indices from \lstinline{0} and Julia from \lstinline{1}.

\begin{figure}
    \centering
    \includegraphics[width=\linewidth]{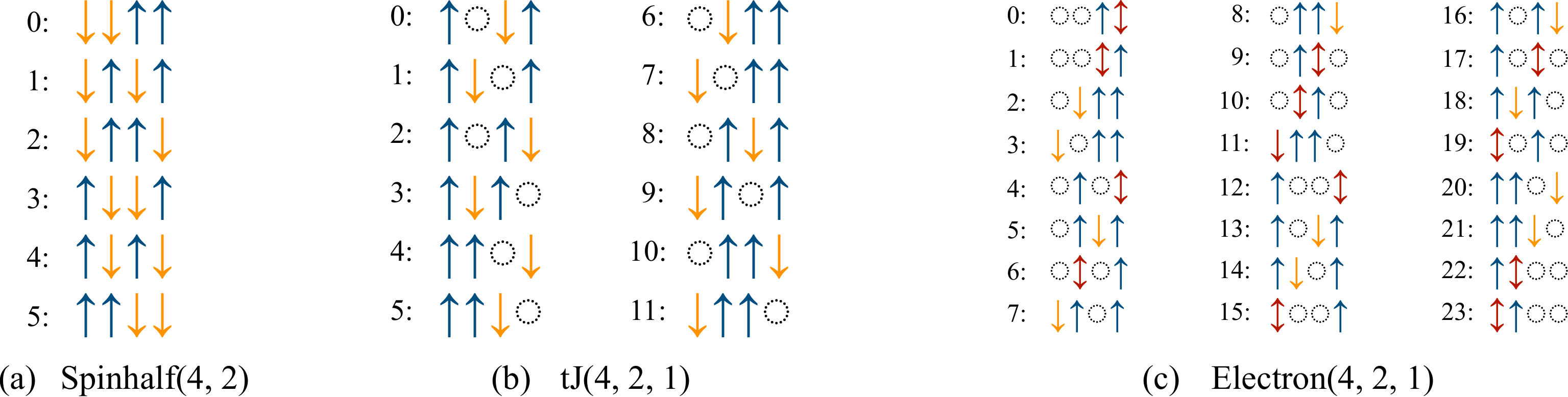}
    \caption{Enumeration of computational basis states of different Hilbert space blocks. (a) \lstinline{Spinhalf} block on $N=4$ sites with $n_{\uparrow}=2$ $\uparrow$-spins. (b) \lstinline{tJ} block on $N=4$ sites with $n_{\uparrow}=2$ $\uparrow$-spins and $n_{\downarrow}=1$ $\downarrow$-spins. (c) \lstinline{Electron} block on $N=4$ sites with $n_{\uparrow}=2$ $\uparrow$-spins and $n_{\downarrow}=1$ $\downarrow$-spins. The index of a given basis configuration (represented as an object of type \lstinline{ProductState}) can be obtained using the \lstinline{index} function.}
    \label{fig:hilbertspaces_nosym}
\end{figure}

The precise output depends on which Hilbert space is chosen. Currently, XDiag features three distinct types of Hilbert spaces:

\begin{enumerate}
    \item \lstinline{Spinhalf}: $S=1/2$ spins; each site is either occupied by an $\uparrow$-spin or a $\downarrow$-spin.
    \item \lstinline{Electron}: spin $S=1/2$ electrons; each site is either empty $\emptyset$, occupied by an $\uparrow$-spin or $\downarrow$-spin electron, or is doubly occupied $\updownarrow$.
    \item \lstinline{tJ}: spin $S=1/2$ electrons without double occupancies; each site is either empty $\emptyset$, occupied by an $\uparrow$-spin or $\downarrow$-spin electron.
\end{enumerate}
Frequently, many-body systems feature certain symmetries and conservation laws. Common conservation laws include particle number, spin, or momentum conservation. The Hilbert space can then be subdivided into blocks, which are labeled by the respective conserved quantities. Blocks of a Hilbert space with a given particle number can be easily created by handing further arguments when constructing the Hilbert space specifying the particle numbers. The number of $\uparrow$-spins in a \lstinline{Spinhalf} block can be specified via,

\noindent
\begin{minipage}[t]{.48\textwidth}
\begin{lstlisting}[language=C++,title=C++,belowcaptionskip=-10pt]
int nup = 2;
auto b1 = Spinhalf(N, nup);

int ndn = 1;
auto b2 = tJ(N, nup, ndn);
auto b3 = Electron(N, nup, ndn);
\end{lstlisting}
\end{minipage}
\hfill
\begin{minipage}[t]{.48\textwidth}
\begin{lstlisting}[language=Julia,title=Julia,belowcaptionskip=-10pt]
nup = 2
b1 = Spinhalf(N, nup)

ndn = 1
b2 = tJ(N, nup, ndn)
b3 = Electron(N, nup, ndn)
\end{lstlisting}
\end{minipage}
The result of printing out the configurations of specific blocks is shown in Fig.~\ref{fig:hilbertspaces_nosym}. This enumeration is important to interpret the coefficients of wave functions. By printing out the basis states of spin configurations, the user can also assess how the computational basis states are ordered internally in XDiag.

\subsection{Operators}
Besides Hilbert spaces, the second key objects in quantum mechanics are operators. In a many-body setting, we consider operators of the form,
\begin{equation}
O = \sum_{A\subseteq \mathcal{L}} c_A O_A,   
\end{equation}
where $O_A$ denotes a local operator acting on sites $A=\{a_1, \ldots, a_{l_A}\}$, $\mathcal{L}$ denotes the lattice, and $c_{A}$ are coupling constants. In the case of the Heisenberg model, we would therefore have $O_{A} = \mathbf{S}_i\cdot\mathbf{S}_j$ and $c_A = J$. In XDiag, the local operators are represented via an \lstinline{Op} object, while the sum is represented by an \lstinline{OpSum} object. These values of the coupling constants $c_A$ can either be a real or complex number or a string that later needs to be replaced. The Hamiltonian of a spin $S=1 / 2$ Heisenberg chain is created in the following way:

\noindent
\begin{lstlisting}[language=C++,title=C++]
auto ops = OpSum();
for (int i=0; i<N; ++i) {
    int s1 = i;
    int s2 = (i+1) % N;
    ops += "J" * Op("SdotS", {s1, s2});
}
ops["J"] = 1.0;
\end{lstlisting}

\begin{lstlisting}[language=Julia,title=Julia]
ops = OpSum()
for i in 1:N
    s1 = i
    s2 = mod1(i+1, N)
    ops += "J" * Op("SdotS", [s1, s2])
end
ops["J"] = 1.0
\end{lstlisting}
We first create an empty \lstinline{OpSum} and then add additional terms to it. The first part of the product denotes the coupling constant, here given as a string. Alternatively, one could have directly used real / complex numbers here. The second part of the product is a single \lstinline{Op} object. It is created with two inputs:
\begin{enumerate}
    \item The type, here \lstinline{SdotS} denoting an operator of the form $\mathbf{S}_{i} \cdot \mathbf{S}_{j}$. XDiag features a wide variety of operator types, see appendix~\ref{app:operator_types} for a detailed list of available options.
    \item An array defining which site the operator lives on. Notice that in Julia we start counting the sites from 1, while in C++ we start counting the sites from 0.
\end{enumerate}

\subsection{Matrix type}

The \lstinline{Matrix} interaction type is a special type with which one can define generic interactions for the \lstinline{Spinhalf} block. In addition to the \lstinline{type} and \lstinline{sites} arguments, a numerical matrix is required when constructing the \lstinline{Op} object. The matrix describes the operator acting on the $ 2^n$-dimensional space spanned by the $n$ sites of the operator. For example, we can represent a $S^x$ spin operator as,

\begin{lstlisting}[language=C++,title=C++]
auto sx = arma::mat({{0, 1},{1, 0}});
auto op = Op("Matrix", 0, sx);
\end{lstlisting}
	
\noindent More generically, we can use this mechanism to construct arbitrary spin interactions, e.g.	

\begin{lstlisting}[language=C++,title=C++]
auto sx = arma::mat({{0, 1},{1, 0}});
auto sz = arma::mat({{0.5, 0},{0, -0.5}});

arma::mat sxsz = arma::kron(sx, sz);
arma::mat sxszsxsz = arma::kron(sxsz, sxsz);
 
auto op_sxsz = Op("Matrix", {0, 1}, sxsz);
auto op_sxszsxsz = Op("Matrix", {0, 1, 2, 3}, sxsz);
\end{lstlisting}

\noindent Here, we have used the Kronecker product function \lstinline{kron}.

\subsection{Complex couplings}
XDiag allows all couplings to be complex. Depending on the operator type, a complex coupling can have two meanings:

\begin{enumerate}
    \item A complex prefactor $c$ which upon hermitian conjugation with \href{hc.md}{hc} gets conjugated to $c^\star$. This is the case for the following interaction types:
    \texttt{HubbardU}, \texttt{Cdagup}, \texttt{Cdagdn}, \texttt{Cup}, \texttt{Cdn}, \texttt{Nup}, \texttt{Ndn}, \texttt{Ntot}, \texttt{NtotNtot}, \texttt{SdotS}, \texttt{SzSz}, \texttt{Sz}, \texttt{S+}, \texttt{S-}, \texttt{ScalarChirality}, \texttt{tJSzSz}, \texttt{tJSdotS}, \texttt{Matrix}.
    Thus, a complex coupling can turn a Hermitian operator to a non-Hermitian operator.

    \item The coupling is part of the definition of the operator. For example, a hopping operator of the form
    $$ ( t c^\dagger_{i\sigma}c_{j\sigma} + \textrm{h.c.}) = ( t c^\dagger_{i\sigma}c_{j\sigma} + t^\star c^\dagger_{j\sigma}c_{i\sigma}) $$
    A complex coupling $t$ gives the hopping a phase, but the overall operator remains Hermitian and, thus, invariant under Hermitian conjugation. This holds for the types \texttt{Hop}, \texttt{Hopup}, \texttt{Hopdn}, \texttt{Exchange}. In the latter case, complex spin exchange \texttt{Exchange} is defined as,
    $$ \frac{1}{2}( J S^+_i S^-_j + J^\star S^-_iS^+_j)$$
\end{enumerate}

\subsection{Matrix representation}\label{sec:matrix-representation}
Given an operator in the form of an \lstinline{OpSum} object and (a sector of) a Hilbert space in the form of a certain Hilbert space block, a dense matrix representation of the operator on the computational basis of the block can be computed using the \lstinline{matrix} function.

\noindent
\begin{minipage}[t]{.48\textwidth}
\begin{lstlisting}[language=C++,title=C++,belowcaptionskip=-10pt]
arma::mat H = matrix(ops, block);
H.print();
\end{lstlisting}
\end{minipage}
\hfill
\begin{minipage}[t]{.48\textwidth}
\begin{lstlisting}[language=Julia,title=Julia,belowcaptionskip=-10pt]
H = matrix(ops, block);
display(H)
\end{lstlisting}
\end{minipage}

\noindent In C++, XDiag is using the Armadillo library~\cite{armadillo} with the \lstinline{arma} namespace. The Armadillo library serves as the linear algebra backend of XDiag and can be used to perform further calculations. To compute all eigenvalues and eigenvectors of a Hamiltonian, i.e., to perform a full exact diagonalization, standard linear algebra routines can be used.

\noindent
\begin{minipage}[t]{.45\textwidth}
\begin{lstlisting}[language=C++,title=C++,belowcaptionskip=-10pt]
arma::vec evals;
arma::mat evecs;
arma::eig_sym(evals, evecs, H);
\end{lstlisting}
\end{minipage}
\hfill
\begin{minipage}[t]{.52\textwidth}
\begin{lstlisting}[language=Julia,title=Julia,belowcaptionskip=-10pt]
(evals, evecs) = eigen(Symmetric(H))
\end{lstlisting}
\end{minipage}

\noindent In Julia, the \lstinline{eigen} and \lstinline{Symmetric} functions are part of the \lstinline{LinearAlgebra} standard library.

\subsection{States}
Quantum states $| \psi \rangle$ are represented in XDiag by using a \lstinline{State} object. A state with zero coefficients is created either implicitly by calling the constructor of \lstinline{State} with a given block, or explicitly by calling the \lstinline{zero_state} function.

\noindent
\begin{minipage}[t]{.49\textwidth}
\begin{lstlisting}[language=C++,title=C++,belowcaptionskip=-10pt]
bool real = true;
auto psi1 = State(b, real);
auto psi2 = zero_state(b, real);
\end{lstlisting}
\end{minipage}
\hfill
\begin{minipage}[t]{.48\textwidth}
\begin{lstlisting}[language=Julia,title=Julia,belowcaptionskip=-10pt]
real = true
psi1 = State(b, real=real)
psi2 = zero_state(b, real=real)
\end{lstlisting}
\end{minipage}

\noindent We hereby create a state with real (double precision) coefficients or complex (double precision) coefficients. The parameter \lstinline{real} is optional, can be omitted and defaults to \lstinline{true}. A state with a given vector of coefficients can also be created.

\noindent
\begin{minipage}[t]{.49\textwidth}
\begin{lstlisting}[language=C++,title=C++,belowcaptionskip=-10pt]
int d = size(block);
arma::vec v(d, arma::fill::randu);
auto psi = State(b, v);
\end{lstlisting}
\end{minipage}
\hfill
\begin{minipage}[t]{.48\textwidth}
\begin{lstlisting}[language=Julia,title=Julia,belowcaptionskip=-10pt]
d = size(block)
v = rand(d)
psi = State(block, v)
\end{lstlisting}
\end{minipage}

\noindent Moreover, we can create product states as well as random states (with normal $\mathcal{N}(0, 1)$ distributed coefficients).

\noindent
\begin{minipage}[t]{.49\textwidth}
\begin{lstlisting}[language=C++,title=C++,belowcaptionskip=-10pt]
auto psi1 = product_state(block,
             {"Up", "Dn"});
auto psi2 = random_state(block);
\end{lstlisting}
\end{minipage}
\hfill
\begin{minipage}[t]{.48\textwidth}
\begin{lstlisting}[language=Julia,title=Julia,belowcaptionskip=-10pt]
psi1 = product_state(block, 
        ["Up", "Dn"])
psi2 = random_state(block)
\end{lstlisting}
\end{minipage}

\noindent The $2$-norm $\parallel  | \psi \rangle\parallel_2$ and dot product $\langle \psi_1 | \psi_2 \rangle$ of states can easily be computed using the \lstinline{norm} and \lstinline{dot}/\lstinline{dotC} functions.

\noindent
\begin{minipage}[t]{.49\textwidth}
\begin{lstlisting}[language=C++,title=C++,belowcaptionskip=-10pt]
double nrm = norm(psi);
double d = dot(psi1, psi2);
complex dc = dotC(psi1, psi2);
\end{lstlisting}
\end{minipage}
\hfill
\begin{minipage}[t]{.48\textwidth}
\begin{lstlisting}[language=Julia,title=Julia,belowcaptionskip=-10pt]
nrm = norm(psi)
d = dot(psi1, psi2)
\end{lstlisting}
\end{minipage}

\noindent The function \lstinline{dotC} is only available in C++, and returns a complex (double precision) number whenever one of the involved states is complex. This is necessary, as the return type of a function must be known at compile time in C++, whereas Julia permits dynamic typing. The coefficients of a given state can be retrieved using the \lstinline{vector} or \lstinline{vectorC} function.

\noindent
\begin{minipage}[t]{.49\textwidth}
\begin{lstlisting}[language=C++,title=C++,belowcaptionskip=-10pt]
arma::vec v = vector(psi);
arma::cx_vec vc = vectorC(psi);
\end{lstlisting}
\end{minipage}
\hfill
\begin{minipage}[t]{.48\textwidth}
\begin{lstlisting}[language=Julia,title=Julia,belowcaptionskip=-10pt]
v = vector(psi)
\end{lstlisting}
\end{minipage}

\noindent Again, the function \lstinline{vectorC} only exists in the C++ version since the return type needs to be known at compile time. In Julia, the type of the vector is decided at runtime. 

Finally, we can apply an operator \lstinline{OpSum} to a state using the \lstinline{apply} function.

\noindent
\begin{minipage}[t]{.49\textwidth}
\begin{lstlisting}[language=C++,title=C++,belowcaptionskip=-10pt]
auto phi = apply(H, psi);
\end{lstlisting}
\end{minipage}
\hfill
\begin{minipage}[t]{.48\textwidth}
\begin{lstlisting}[language=Julia,title=Julia,belowcaptionskip=-10pt]
phi = apply(H, psi)
\end{lstlisting}
\end{minipage}

\noindent Importantly, if the block of the \lstinline{State}  object has a well-defined quantum number, for example, a conserved particle number, XDiag will automatically detect the quantum number of the resulting state or report an error if the operator does not have a well-defined quantum number. This could be the case, for example, when applying a raising or lowering operator on a particle number conserving state. The \lstinline{apply} function acts on a state without creating a matrix representation of the operator, sometimes referred to as \textit{on-the-fly} matrix application.

\subsection{Iterative algorithms}
XDiag features several iterative algorithms to perform diagonalization and time evolution, which do not require dense matrix storage of the involved operators. Instead, applications of operators are implemented on-the-fly (i.e., matrix-free) to minimize memory requirements.

\subsubsection{Diagonalization}
A fundamental property of a quantum system is its ground state energy, which in XDiag can be easily computed using the \lstinline{eigval0} function.

\noindent
\begin{minipage}[t]{.54\textwidth}
\begin{lstlisting}[language=C++,title=C++,belowcaptionskip=-10pt]
double e0 = eigval0(H, block);
\end{lstlisting}
\end{minipage}
\hfill
\begin{minipage}[t]{.43\textwidth}
\begin{lstlisting}[language=Julia,title=Julia,belowcaptionskip=-10pt]
e0 = eigval0(H, block)
\end{lstlisting}
\end{minipage}

\noindent Similarly, the ground state can be computed using the \lstinline{eig0} function.

\noindent
\begin{minipage}[t]{.54\textwidth}
\begin{lstlisting}[language=C++,title=C++,belowcaptionskip=-10pt]
auto [e0 , psi0] = eig0(H, block);
\end{lstlisting}
\end{minipage}
\hfill
\begin{minipage}[t]{.43\textwidth}
\begin{lstlisting}[language=Julia,title=Julia,belowcaptionskip=-10pt]
e0, psi0 = eig0(H, block)
\end{lstlisting}
\end{minipage}

\noindent Here, \lstinline{e0} is a double precision real number and \lstinline{psi0} is a \lstinline{State} object. We employ the Lanczos algorithm~\cite{Lanczos1950} to perform iterative diagonalizations. While the functions \lstinline{eigval0} and \lstinline{eig0} are convenient, XDiag offers also more refined routines called \lstinline{eigvals_lanczos} and \lstinline{eigs_lanczos} which can target higher excited states and offer more control over the convergence properties. Moreover, they return the convergence criterion as well as the tridiagonal matrix in the Lanczos algorithm, which contains more information than only the extremal eigenvalues. The functions \lstinline{eig0} and \lstinline{eigs_lanczos} for computing eigenvalues perform the Lanczos iteration twice, first to compute the tridiagonal matrix and in a second run to build the corresponding eigenvectors in order to minimize memory requirements. For a precise description of these methods, we refer to the documentation~\cite{documentation}.

\subsubsection{Time evolution}
Besides diagonalization, XDiag also offers iterative algorithms to perform (imaginary-) time evolutions of the form,
\begin{equation}
    |\phi(t)\rangle = e^{-iHt} |\psi_0\rangle \quad \text{or} \quad |\eta(t)\rangle = e^{-\tau H} |\psi_0\rangle.
\end{equation}

\noindent
\begin{minipage}[t]{.54\textwidth}
\begin{lstlisting}[language=C++,title=C++,belowcaptionskip=-10pt]
double t = 1.0;
auto phi = time_evolve(H, psi0, t);
\end{lstlisting}
\end{minipage}
\hfill
\begin{minipage}[t]{.43\textwidth}
\begin{lstlisting}[language=Julia,title=Julia,belowcaptionskip=-10pt]
t = 1.0
phi = time_evolve(H, psi0, t)
\end{lstlisting}
\end{minipage}

\noindent The time evolution can be performed by two distinct algorithms. The first is the memory-efficient Lanczos algorithm, described in Ref.~\cite{Hochbruck1997}, which runs a Lanczos algorithm twice, to first compute the tridiagonal matrix and then build the time-evolved state. The second is an efficient algorithm proposed in \textit{Expokit}~\cite{Sidje1998}. While this algorithm is computationally more efficient and highly accurate, it has higher memory requirements. The algorithm employed can be set using the optional \lstinline{algorithm} argument, which by default is set to \lstinline{lanczos} using the memory-efficient algorithm of Ref.~\cite{Hochbruck1997}. Also, more direct control of both algorithms is provided by the functions \lstinline{evolve_lanczos} and \lstinline{time_evolve_expokit}, which allow more control and return further data such as the tridiagonal matrix of the Lanczos algorithm or error estimates, respectively. We refer to the documentation~\cite{documentation} for a more detailed description of the parameters and return types of these functions.

\subsection{Measurements}
Measurements in the form of expectation values of wavefunctions,
\begin{equation}
   \langle O\rangle =  \langle \psi | O | \psi \rangle, 
\end{equation}
can be evaluated using the \lstinline{inner} function. For example, we compute a static spin correlation $\langle S_{0}^{z} S_{j}^{z}\rangle$ between site $0$ (resp. $1$ in Julia) and $j$.

\noindent
\begin{minipage}[t]{.54\textwidth}
\begin{lstlisting}[language=C++,title=C++,belowcaptionskip=-10pt]
for (int i=0; i<N; ++i) {
    auto op = Op("SzSz", {0, i});
    double corr = inner(op, psi0);
}
\end{lstlisting}
\end{minipage}
\hfill
\begin{minipage}[t]{.43\textwidth}
\begin{lstlisting}[language=Julia,title=Julia,belowcaptionskip=-10pt]
for i in 1:N
    op = Op("SzSz", {1, i})
    corr = inner(op, psi0)
end
\end{lstlisting}
\end{minipage}

\noindent Notice, that in Julia sites start counting from $1$, whereas in C++ sites are counted from $0$. Furthermore, if a complex wave function or operator is involved, the function \lstinline{innerC} in C++ should be called, which returns a complex number. In Julia only \lstinline{inner} is available whose return type is decided at runtime. 

\subsection{Input / Output}
Julia features a variety of packages facilitating input and output of data. For C++, XDiag provides convenient functionality for TOML and HDF5 files.

\subsubsection{Reading from TOML}
While XDiag allows for defining a Hamiltonian or other operators in code, it is often preferable to define objects in advance in an input file. In XDiag, we use the TOML language to define basic objects in simple files. Among those, operators represented as an \lstinline{OpSum} can be specified in a simple format. As an example, the Hamiltonian of the $N=8$ site Heisenberg chain we created above can be defined in the following way.

\begin{lstlisting}[language=Toml,title=spinhalf\_chain.toml]
Interactions = [
    ["J", "SdotS", 0, 1],
    ["J", "SdotS", 1, 2],
    ["J", "SdotS", 2, 3],
    ["J", "SdotS", 3, 4],
    ["J", "SdotS", 4, 5],
    ["J", "SdotS", 5, 6],
    ["J", "SdotS", 6, 7],
    ["J", "SdotS", 7, 0]
]
\end{lstlisting}

\noindent The first entry in every list element denotes the coupling constant $J$, the second entry denotes the type \lstinline{SdotS}, and the following two entries are the sites of the operator. To read in such a Hamiltonian from a TOML file, we can use the \lstinline{FileToml} object together with the \lstinline{read_opsum} function.

\begin{lstlisting}[language=C++,title=C++]
auto fl = FileToml("spinhalf_chain.toml");
auto ops = read_opsum(fl, "Interactions");
ops["J"] = 1.0;
\end{lstlisting}

\begin{lstlisting}[language=Julia,title=Julia]
fl = FileToml("spinhalf_chain.toml")
ops = read_opsum(fl, "Interactions")
ops["J"] = 1.0
\end{lstlisting}

\noindent In the TOML file, the coupling constant $J$ is defined as a string, which is then set to a numerical value in the application code. Alternatively, the coupling constant could also be explicitly defined by a numerical value in the TOML file. Complex numbers $x + iy$ can be represented by a size-two array of the form \lstinline{[x, y]}. XDiag also features the functions \lstinline{read_permutation_group} and \lstinline{read_representation} to conveniently read \lstinline{PermutationGroup} and \lstinline{Representation} objects used to describe symmetries from a file; see section~\ref{sec:symmetrytoml}.

\noindent The TOML input files assume that counting starts at $0$ for both the C++ and Julia version, such that the same input file can be used for both languages. In Julia, however, the site index is increased by one after reading in the \lstinline{OpSum}.

\subsubsection{Writing results to HDF5}
The results of numerical simulations are typically stored in output files. A standard scientific data format is the HDF5 format. Julia supports input and output to HDF5 with the \lstinline{HDF5.jl} package. For C++ we provide a convenient way of writing results to HDF5 files. In general all numerical data, including scalar real/complex numbers as well as Armadillo vectors and matrices can be easily written to file using the \lstinline{FileH5} object.

\begin{lstlisting}[language=C++,title=C++]
auto f1 = FileH5("output.h5", "w!");
f1["e0"] = e0;
f1["evals"] = evals;
f1["evecs"] = evecs;
\end{lstlisting}

\noindent The second argument \lstinline{"w!"} specifies the access mode to the file. \lstinline{"w!"} is the forced write mode, where preexisting files are overwritten. Additionally, a protected write mode \lstinline{"w"} raising an error for existing files, and a read mode \lstinline{"r"} are provided.

\subsection{Symmetries}
\label{subsec:symmetries}

Symmetries are fundamental properties of any physical system. In quantum mechanics, symmetries of a Hamiltonian lead to a set of conserved quantities, also referred to as quantum numbers. Mathematically, quantum numbers label irreducible representations of symmetry groups. A particularly important set of symmetries are space group symmetries, such as translation symmetries in a solid or point group symmetries in molecules. Abstractly, these symmetries are permutations of sites of the interaction graph, which can be efficiently made use of in ED simulations, see e.g. ~\cite{Schaefer2020}. Permutation groups then have irreducible representations, which can denote the momentum for translation groups or angular momentum in point groups. XDiag can employ these symmetries, allowing for more efficient computations as well as physical insights, e.g., via tower-of-states analyses~\cite{Wietek2016b}. Internally, XDiag checks for consistency of symmetries and representations and allows the use of permutational symmetries also for multi-site interactions.

\subsubsection{Permutations}
Mathematically, a permutation $\pi$ of $N$ elements is a bijective mapping,
\begin{equation}
\pi : \{ 1, 2, \ldots, N\} \rightarrow  \{ 1, 2, \ldots, N\},
\end{equation}
where every integer in the range from $1$ to $N$ is mapped to a distinct number from $1$ to $N$. Such a mapping is usually written as, 
\begin{equation}
    \pi = \begin{pmatrix} 
    1 & 2& \ldots & N \\
    \pi(1) & \pi(2) & \ldots & \pi(N) \\
    \end{pmatrix}.
\end{equation}
For example, a translation operator $T$ on a chain with periodic boundary conditions with $N$ sites can be written as,
\begin{equation}
    T = \begin{pmatrix} 
    1 & 2& \ldots & N-1 & N \\
    2 & 3& \ldots & N & 1 \\
    \end{pmatrix}.
\end{equation}
In XDiag, such a mapping is represented by a \lstinline{Permutation} object, which is created by the integer vector of permuted indices.

\noindent
\begin{minipage}[t]{.54\textwidth}
\begin{lstlisting}[language=C++,title=C++,belowcaptionskip=-10pt]
auto T = Permutation({1, 2, 3, 4, 
                      5, 6, 7, 0});
\end{lstlisting}
\end{minipage}
\hfill
\begin{minipage}[t]{.43\textwidth}
\begin{lstlisting}[language=Julia,title=Julia,belowcaptionskip=-10pt]
T = Permutation([2, 3, 4, 5, 
                 6, 7, 8, 1])
\end{lstlisting}
\end{minipage}

\noindent Notice that also here we start counting from 1 in Julia, and from 0 in C++. \lstinline{Permutation} objects can be multiplied (i.e., concatenated), inverted, and raised to a power. A set of permutations can form a mathematical group. To define a mathematical permutation group, we can construct a \lstinline{PermutationGroup} object by handing a vector of \lstinline{Permutation} objects.

\begin{lstlisting}[language=C++,title=C++]
auto group = PermutationGroup({
    pow(T, 0), pow(T, 1), pow(T, 2), pow(T, 3),
    pow(T, 4), pow(T, 5), pow(T, 6), pow(T, 7)});
\end{lstlisting}

\begin{lstlisting}[language=Julia,title=Julia]
group = PermutationGroup([T^0, T^1, T^2, T^3, T^4, T^5, T^6, T^7])
\end{lstlisting}

\noindent Internally, the group axioms are validated, and an error will be thrown if not all group properties are fulfilled. This means that the existence of the identity permutation is required. Every permutation also necessitates its inverse to be present, and the product of two permutations needs to be defined as well.

\subsubsection{Representations}\label{subsubsec:representations}
Mathematically, a representation $\rho$ of a group $G$ is a mapping from a group to the group of invertible matrices $\textrm{GL}(V)$ on a vector space $V$, 
\begin{equation}
    \rho: G \mapsto \textrm{GL}(V),
\end{equation}
fulfilling the group homomorphism property,
\begin{equation}
\rho(g \cdot h) = \rho(g) \cdot \rho(h) \quad \textrm{for all} \quad g, h \in G.
\end{equation}
The dimension of $V$, i.e., the dimension of the matrices, is referred to as the dimension of the representation. The character $\chi_\rho$ of a representation $\rho$ denotes the trace of the representation matrices, 
\begin{equation}
    \chi_\rho(g) = \textrm{Tr}[\rho(g)] \in \mathbf{C}.
\end{equation}
One-dimensional representations play an important role in the theory of space groups. We identify the representation matrices of one-dimensional representations with their characters. 

An important example is the cyclic group $C_N$ of order $N$ consisting e.g. of translations $\{ \textrm{I}, T, T^2, \ldots, T^{(N-1)}\}$, which has $N$ irreducible one-dimensional representations $\rho_k$ with associated character functions $\chi_k$. Those can be labeled by numbers $k = 2\pi l/N$ for $l \in \{1, \ldots, N\}$, such that the characters of the $k$-th irreducible representation (irrep) are, 
\begin{equation}
    \chi_k(T^n) = e^{i k n}.
\end{equation}
Hence, each irrep corresponds to a certain momentum $k$ in physical terms. In XDiag, a one-dimensional representation is represented by the \lstinline{Representation} object and can be created by handing a \lstinline{PermutationGroup} and the list of characters.

\noindent
\begin{minipage}[t]{.54\textwidth}
\begin{lstlisting}[language=C++,title=C++,belowcaptionskip=-10pt]
auto chi = arma::vec(
       {1.0, -1.0, 1.0, -1.0,
        1.0, -1.0, 1.0, -1.0});
auto k = Representation(group, chi);
\end{lstlisting}
\end{minipage}
\hfill
\begin{minipage}[t]{.43\textwidth}
\begin{lstlisting}[language=Julia,title=Julia,belowcaptionskip=-10pt]
chi = [1.0, -1.0, 1.0, -1.0, 
       1.0, -1.0, 1.0, -1.0]
k = Representation(group, chi)
\end{lstlisting}
\end{minipage}

\noindent Upon creation of a representation, XDiag verifies whether the group axioms as well as the homomorphism property of the characters are fulfilled, i.e.,
\begin{equation}
f * g=h \Rightarrow \chi(f) \cdot \chi(g)=\chi(h).
\end{equation}

\subsection{Symmetry-adapted blocks}
Representations $\rho$ of a permutation symmetry group $G$ can then be used to create symmetry-adapted blocks. Given a computational basis state $|\mathbf{\sigma}\rangle = |\sigma_1\sigma_2\cdots\sigma_N\rangle$, the corresponding symmetry-adapted state is given as,
\begin{equation}
    |\mathbf{\sigma}_{\rho}\rangle \equiv \frac{1}{N_{\rho,\psi}} \sum_{g \in G} \chi_{\rho}(g)^{*} g |\mathbf{\sigma}\rangle.
\end{equation}
Here, the action of a permutation $g$ on a product state $|\sigma\rangle$ is defined as,
\begin{equation}
    g|\mathbf{\sigma}\rangle = g|\sigma_1\sigma_2\cdots\sigma_N\rangle = 
    |\sigma_{g(1)}\sigma_{g(2)}\cdots\sigma_{g(N)}\rangle,
\end{equation}
and $N_{\rho,\psi}$ denotes the normalization constant.
For example, consider a spin $S=1/2$ system on a four-site chain with a fourfold translation symmetry group $C_4$, the computational basis state $|\mathbf{\sigma}\rangle = |\downarrow\downarrow\uparrow\uparrow\rangle$ and the irreducible representation with $k=\pi$. Then the symmetry-adapted basis state of $|\mathbf{\sigma}\rangle$ is given as,
\begin{equation}
   |\mathbf{\sigma}_\pi\rangle = \frac{1}{2}\left( 
   |\downarrow\downarrow\uparrow\uparrow\rangle
   -|\downarrow\uparrow\uparrow\downarrow\rangle
   +|\uparrow\uparrow\downarrow\downarrow\rangle
   -|\uparrow\downarrow\downarrow\uparrow\rangle
   \right).
\end{equation}
We refer to the states $\{g|\mathbf{\sigma}\rangle \; | \; g \in G\}$ as the orbit of $|\mathbf{\sigma}\rangle$.
XDiag represents each basis state by an integer value. To represent a symmetry-adapted basis state, XDiag uses the state in the orbit of $|\mathbf{\sigma}\rangle$ with the minimal associated integer value, which is then called the \textit{representative} of the orbit. For more details on symmetry-adapted wavefunctions in the context of exact diagonalization we refer to Refs.~\cite{Weisse2008,WietekThesis}.

We can conveniently create a symmetry-adapted block in XDiag by handing a \lstinline{Representation} object to the constructor of a block.

\noindent
\begin{minipage}[t]{.54\textwidth}
\begin{lstlisting}[language=C++,title=C++,belowcaptionskip=-10pt]
auto blk = Spinhalf(N, nup, irrep);
\end{lstlisting}
\end{minipage}
\hfill
\begin{minipage}[t]{.43\textwidth}
\begin{lstlisting}[language=Julia,title=Julia,belowcaptionskip=-10pt]
blk = Spinhalf(N, nup, irrep)
\end{lstlisting}
\end{minipage}

\noindent A symmetry-adapted block shares most functionality with conventional blocks, e.g., we can iterate over the (representatives of the symmetry-adapted) basis states.

\noindent
\begin{minipage}[t]{.48\textwidth}
\begin{lstlisting}[language=C++,title=C++,belowcaptionskip=-10pt]
for (auto spins: blk) {
    Log("{}", to_string(spins));
}
\end{lstlisting}
\end{minipage}
\hfill
\begin{minipage}[t]{.48\textwidth}
\begin{lstlisting}[language=Julia,title=Julia,belowcaptionskip=-10pt]
for spins in blk
    println(to_string(spins))
end
\end{lstlisting}
\end{minipage}

\begin{figure}
    \centering
    \includegraphics[width=\linewidth]{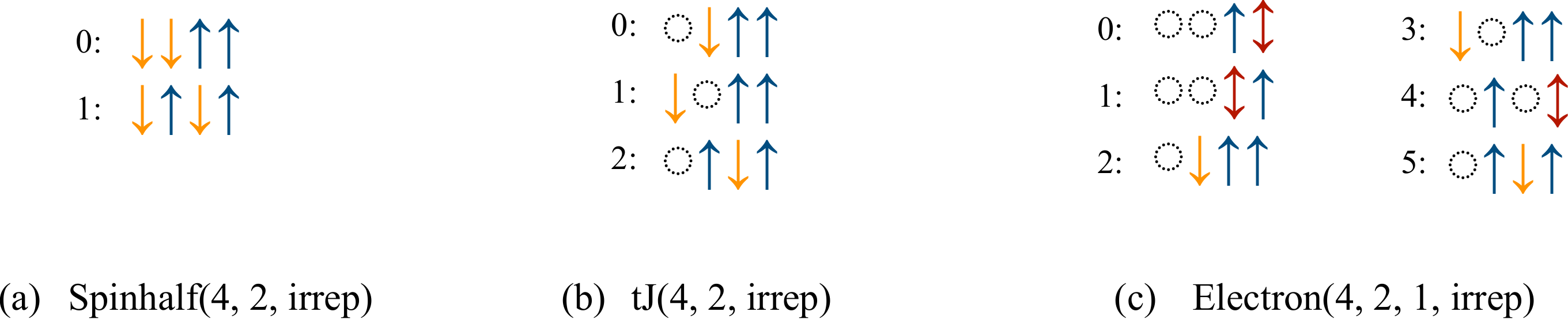}
    \caption{Enumeration of computational basis states of symmetry-adapted Hilbert space blocks with translational symmetry in the $k=0$ representation. (a) \lstinline{Spinhalf} block on $N=4$ sites with $n_{\uparrow}=2$ $\uparrow$-spins. (b) \lstinline{tJ} block on $N=4$ sites with $n_{\uparrow}=2$ $\uparrow$-spins and $n_{\downarrow}=1$ $\downarrow$-spins. (c) \lstinline{Electron} block on $N=4$ sites with $n_{\uparrow}=2$ $\uparrow$-spins and $n_{\downarrow}=1$ $\downarrow$-spins.}
    \label{fig:hilbertspaces_sym}
\end{figure}

\noindent A sample output of symmetry-adapted blocks of different kinds is shown in Fig.~\ref{fig:hilbertspaces_sym}. Symmetry-adapted blocks can, among other use cases, be used to create a representation of an operator in the symmetry-adapted basis using the \lstinline{matrix} function and create symmetric states by constructing \lstinline{State} objects. Symmetry-adapted operators, e.g., created using \lstinline{symmetrize} as described in section~\ref{sec:symmetrize}, in the form of an \lstinline{OpSum} can also be applied to symmetry-adapted states, where consistency of quantum numbers is validated.

\subsection{Reading symmetries from TOML files}
\label{sec:symmetrytoml}
Permutation groups and representations can be defined in a TOML file and read in. A typical input TOML file defining a symmetry group and corresponding representations is shown below.

\begin{lstlisting}[language=TOML,title=symmetries.toml]
Symmetries = [
  [0, 1, 2, 3],
  [3, 0, 1, 2],
  [2, 3, 0, 1],
  [1, 2, 3, 0],
]

# Irreducible representations
[k.zero]
characters = [
  [1.0, 0.0],
  [1.0, 0.0],
  [1.0, 0.0],
  [1.0, 0.0],
]

[k.pihalf]
characters = [
  [1.0, 0.0],
  [0.0, 1.0],
  [-1.0, 0.0],
  [0.0, -1.0],
]

[k.pi]
characters = [
  [1.0, 0.0],
  [-1.0, 0.0],
  [1.0, 0.0],
  [-1.0, 0.0],
]
\end{lstlisting}

\noindent The permutation group is defined as an integer matrix, whose rows correspond to individual permutations. Notice, that we start counting sites from $0$ in this case. This holds regardless of whether the C++ version or the Julia version is used. The input TOML files always start counting at $0$, in Julia, the symmetry number array is increased by one. The representations are given names, \lstinline{k.zero}, \lstinline{k.pihalf}, \lstinline{k.pi}, and have an associated field \lstinline{characters}, which is a vector of complex numbers. Complex numbers are represented by a two-element array. Technically, real representations are also allowed to have one single real number instead. 

The symmetry group and the irreducible representations are then easily read in using the functions \lstinline{read_permutation_group} and \lstinline{read_representation}.


\begin{lstlisting}[language=C++,title=C++]
auto fl = FileToml("symmetries.toml");
auto group = read_permutation_group(fl, "Symmetries");
auto irrep = read_representation(fl, "k.zero", "Symmetries");
\end{lstlisting}

\begin{lstlisting}[language=Julia,title=Julia]
fl = FileToml("symmetries.toml")
group = read_permutation_group(fl, "Symmetries")
irrep = read_representation(fl, "k.zero", "Symmetries")
\end{lstlisting}

\noindent Notice that the \lstinline{read_representation} function is not just handed the name of the representation but also the tag that defines the permutation symmetry group, as the representation needs to know which group it is representing. 

\subsection{Symmetrized operators}
\label{sec:symmetrize}
A useful feature to work with permutation symmetries is to symmetrize a non-symmetric operator. Symmetrization in this context means the following. In general, we are given an operator of the form,
\begin{equation}
O = \sum_{A\subseteq \mathcal{L}} O_A,   
\end{equation}
where $O_A$ denotes a local operator acting on sites $A=\{a_1, \ldots, a_{l_A}\}$ and $\mathcal{L}$ denotes the lattice.
The symmetrization of the operator $O$ with respect to a permutation group $\mathcal{G} = \{\pi_1, \ldots, \pi_M\}$ is then defined as
\begin{equation}
 O^\mathcal{G} = \frac{1}{M}\sum_{A\subseteq \mathcal{L}} \sum_{\pi \in \mathcal{G}}  O_{\pi(A)},
\end{equation}
where $\pi(A) = \{\pi(a_1), \ldots,\pi(a_{l_A})\}$ denotes the permutated set of sites of the local operator $O_A$. Additionally, we can also symmetrize with respect to a representation $\rho$,
\begin{equation}
\label{eq:symmetrizeirrep}
 O^\mathcal{G, \rho} = \frac{1}{M}\sum_{A\subseteq \mathcal{L}} \sum_{\pi \in \mathcal{G}} \chi_\rho(\pi) O_{\pi(A)},
\end{equation}
where $\chi_\rho$ denotes the characters of the representation $\rho$. Such symmetrizations of \lstinline{OpSum} objects can be performed using the \lstinline{symmetrize} function.

\noindent
\begin{minipage}[t]{.48\textwidth}
\begin{lstlisting}[language=C++,title=C++,belowcaptionskip=-10pt]
auto og = symmetrize(ops, group);
auto oi = symmetrize(ops, irrep);
\end{lstlisting}
\end{minipage}
\hfill
\begin{minipage}[t]{.48\textwidth}
\begin{lstlisting}[language=Julia,title=Julia,belowcaptionskip=-10pt]
og = symmetrize(ops, group)
oi = symmetrize(ops, irrep)
\end{lstlisting}
\end{minipage}

\noindent A common use case of symmetrizing operators is to evaluate expectation values of non-symmetric operators on a symmetric \lstinline{State}. For example, we might be interested in the ground state of a model with translation symmetry and then evaluate two-point correlation functions. The ground state is then going to be defined on a symmetric block, but the two-point correlation function is not symmetric with respect to the translation symmetry group. However, the symmetrized two-point correlation function is, in fact, symmetric, and its expectation value on a symmetric state will evaluate to the same number. This is a typical use case of symmetrizing an operator with respect to a group. Another use case of the \lstinline{symmetrize} function is to create operators at a certain momentum, e.g.
\begin{equation}
  S^z(\mathbf{q}) = \frac{1}{M} \sum_{\bf r} e^{i\mathbf{q}\cdot\mathbf{r}} S^z_{\mathbf{r}}.
\end{equation}
Such an operator is of the form in Eq.~\ref{eq:symmetrizeirrep} and can, hence, be easily created using the \lstinline{symmetrize} function with the irrep describing the momentum $\mathbf{q}$. We provide several examples along XDiag demonstrating this functionality and use cases.

\subsection{Sparse Matrix Capabilities}
Working with many-body quantum systems often involves matrices which only have a small number of non-zero elements, also known as \emph{sparse matrices}.
Since storing and handling such objects in the conventional way (i.e. element by element) is very inefficient, XDiag includes basic implementations of three common sparse-matrix types: the coordinate (COO), the compressed-sparse-row (CSR), and the compressed-sparse-column (CSC) formats.

Just as the \lstinline{matrix} function can be used to obtain the full matrix representing a given operator \lstinline{ops} (in the form of an \lstinline{OpSum} object) on a given Hilbert space \lstinline{block} (also see Sec.~\ref{sec:matrix-representation}), there are functions \lstinline{coo_matrix}, \lstinline{csr_matrix}, and \lstinline{csc_matrix} to obtain the same matrix in the respective sparse format.

\begin{lstlisting}[language=C++,title=C++]
auto coo_mat = coo_matrix(ops, block);
auto csr_mat = csr_matrix(ops, block);
auto csc_mat = csc_matrix(ops, block);
\end{lstlisting}

\begin{lstlisting}[language=Julia,title=Julia]
coo_mat = coo_matrix(ops, block)
csr_mat = csr_matrix(ops, block)
csc_mat = csc_matrix(ops, block)
\end{lstlisting}
Note that the C++ implementation distinguishes between real and complex matrices, e.g., there are the \lstinline{csr_matrix} and the \lstinline{csr_matrixC} functions.
The objects returned by these functions are ``raw'' in the sense that they are not an instance of a sparse-matrix implementation of another library, but contain all the information to call the respective sparse-matrix constructor of your sparse-matrix library of choice.
For instance, the output of \lstinline{csc_matrix} can be used to construct the \lstinline{sp_mat} type implemented by the Armadillo library in C++ or the \lstinline{SparseMatrixCSC} type implemented by \lstinline{SparseArrays} (part of the standard library) in Julia.

\begin{lstlisting}[language=C++,title=C++]
auto colptr = arma::conv_to<arma::uvec>::from(csc_mat.colptr);
auto row = arma::conv_to<arma::uvec>::from(csc_mat.row);
auto A = arma::sp_mat(colptr, row, 
                      csc_mat.data, csc_mat.nrows, csc_mat.ncols);
\end{lstlisting}

\begin{lstlisting}[language=Julia,title=Julia]
using SparseArrays
jmat = SparseMatrixCSC(csc_mat.nrows, csc_mat.ncols, 
                       csc_mat.colptr, csc_mat.row, csc_mat.data)
\end{lstlisting}

While the COO, CSR, and CSC formats are supported for extracting sparse matrices from XDiag, only the CSR format is used internally because it is the only one suitable for parallelized matrix-vector multiplications.
The user guide of our online documentation~\cite{documentation} compares the default (matrix-free, i.e., on-the-fly) implementation of Lanczos iterations to using the CSR-based version after building and converting the full Hamiltonian into the CSR format.
Based on these examples, at medium system sizes the runtime of both approaches is comparable, while the matrix-free variant will always require substantially less memory.
Note that the computational effort and memory required to construct the CSR representation grows significantly with system size such that the matrix-free default implementation will generally run faster on larger systems.
However, a clear advantage of sparse-matrix implementations can be expected when sufficient memory is available for the CSR matrix to be pre-computed once before it needs to be reused frequently.

\subsection{Distributed computing}
The standard XDiag library features shared memory parallelization using OpenMP for both the C++ and Julia version. However, we also provide distributed memory parallelization in a separate C++-only library that needs to be compiled independently. To do so, we have to use the flag \lstinline{XDIAG_DISTRIBUTED} when setting up the compilation using \lstinline{CMake}.
\begin{lstlisting}[language=bash]
cmake -S . -B build -D XDIAG_DISTRIBUTED=On
cmake --build build
cmake --install build
\end{lstlisting}
\noindent This will create a distinct \lstinline{xdiag_distributed} library, which is different from the standard \lstinline{xdiag} library. To link an application code to the distributed library, we can use the following \lstinline{CMakeLists.txt} file.
\begin{lstlisting}[language=bash]
cmake_minimum_required(VERSION 3.19)
project(hello_world)
find_package(xdiag_distributed REQUIRED HINTS "/path/to/where/xdiag/should/be/install")
add_executable(main main.cpp)
target_link_libraries(main PRIVATE xdiag::xdiag_distributed)  
\end{lstlisting}
\noindent Notice that this only differs from the original \lstinline{CMakeLists.txt} file shown in Sec.~\ref{sec:applicationCompilation} in two places, where we specify that the \lstinline{xdiag_distributed} library instead of the standard \lstinline{xdiag} library is used. The distributed memory library is built on top of the message passing interface (MPI). Every MPI application needs to initialize and finalize the MPI environment explicitly. Hence, a typical main routine for using the distributed XDiag library with MPI should look similar to the following listing.
\begin{lstlisting}[language=C++,title=C++]
#include <xdiag/all.hpp>
using namespace xdiag;
int main(int argc, char* argv[]) try {
    MPI_Init(argc, argv);
    // genuine XDiag code here
    MPI_Finalize();
} catch (Error e) {
    error_trace(e);
}
\end{lstlisting}

\noindent The functionality described in the previous sections is also available to some degree for the distributed library. Importantly, to use the distributed capabilities we have to change the types of blocks from the standard \lstinline{Spinhalf}, \lstinline{tJ}, and \lstinline{Electron} blocks to the \lstinline{SpinhalfDistributed}, \lstinline{tJDistributed}, and \lstinline{ElectronDistributed} blocks. An example to compute the ground state of a Heisenberg chain using the distributed capabilities is given by,

\begin{lstlisting}[language=C++,title=C++]
auto block = SpinhalfDistributed(N, nup);
OpSum ops;    
for (int i = 0; i < N; ++i) {
    ops += Op("SdotS", {i, (i + 1) % N});
}
double e0 = eigval0(ops, block);
\end{lstlisting}

\noindent At present, one important difference between the distributed and standard library is that the distributed blocks do not yet support symmetrized blocks. This feature will be added in future versions of XDiag.

\section{Examples}
\label{sec:examples}
While we presented the core functionality in section~\ref{sec:usage_guide}, we now provide several examples along with XDiag to demonstrate how we can build more advanced applications from these basic building blocks. The examples are provided in the directory \lstinline{examples} in the C++ source code directory and can be compiled using CMake.

\begin{lstlisting}[language=bash]
cmake -S . -B build -D BUILD_EXAMPLES=On
cmake --build build
\end{lstlisting}

\noindent Most examples are also provided as a Julia script, and visualization scripts for the results are available. A detailed explanation of the provided examples is found in the documentation~\cite{documentation}. Here, we list the provided examples to showcase the versatility of XDiag. 

\subsection{Ground states}
\paragraph{Ground state energy} We compute the ground state energy of a spin $S=1/2$ Heisenberg chain as in Eq.~\ref{eq:heisenberg_model} without using translational symmetries. This minimal example shows how to set up a model and run a simple Lanczos algorithm. 

\paragraph{Ground state correlators} This example demonstrates how ground state expectation values can be computed for the spin $S=1/2$ Heisenberg chain as in Eq.~\ref{eq:heisenberg_model} with periodic boundary conditions. The Hamiltonian is invariant under translations and decomposes into $N$ irreps labeled by lattice momentum $2\pi k/N$, as explained in section~\ref{subsubsec:representations}. The Lanczos algorithm is run on all momentum sub-blocks. Correlation functions are computed by first symmetrizing operators with respect to the symmetry group $C_N$.

\paragraph{Entanglement entropy} We compute the entanglement entropy of the ground state of the spin $S=1/2$ XXZ chain, described by the Hamiltonian
\begin{equation}
\mathcal{H} = J \sum_{n=0}^{N-1} \left(S^x_n S^x_{n+1} + S^y_n S^y_{n+1} + \Delta S^z_n S^z_{n+1}\right). 
\end{equation}
The ground state $|\psi_0\rangle$ is obtained using the Lanczos algorithm for a given value of $\Delta$, and functions are provided to construct the reduced density matrix related to the region with the first $\ell$ spins. This is obtained by tracing out the complementary degrees of freedom,
\begin{equation}
\rho_{\ell} = \text{Tr}_{\bar{\ell}} \left(|\psi_0\rangle\langle\psi_0| \right),    
\end{equation}
which in practice is done by looping over all possible spin configurations of the two regions, as described in subsection~\ref{subsec:hilbert_space}. The entanglement entropy is obtained through the eigenvalues of the reduced density matrix, $\lambda_\alpha$,
\begin{equation}
S_\ell = -\text{Tr}\left( \rho_\ell \ln \rho_\ell  \right)=-\sum_\alpha  \lambda_\alpha \ln \lambda_\alpha.
\end{equation}

\paragraph{Kitaev-Heisenberg honeycomb model}
We consider the Kitaev-Heisenberg model on a honeycomb lattice. The Hamiltonian is given by
\begin{equation}
\mathcal{H} = -K\sum_{\langle i,j\rangle \parallel \tilde{\gamma}} S_i^\gamma S_j^\gamma+J\sum_{\langle i,j\rangle} \boldsymbol{S}_i \cdot \boldsymbol{S}_j,
\end{equation}
where $\boldsymbol{S}_i$ are spin $S=1/2$ operators, $\langle \dots \rangle$ denotes nearest-neighbor sites, and $\gamma=x,y,z$ for bonds in directions $\tilde{\gamma}=(0,1),(\sqrt{3},-1)/2,(-\sqrt{3},-1)/2$, respectively. This model hosts Néel, stripy, zigzag, and ferromagnetic magnetic orders as well as the Kitaev spin liquid phase, depending on the parameter regimes. 

\paragraph{Charge density waves in attractive Hubbard model}
We compute charge correlations in the ground state of the attractive 2D Hubbard model, at half-filling, to show the $(\pi,\pi)$-charge density wave.

\subsection{Full exact diagonalization}
\paragraph{Specific heat, random $t$-$J$ model} We use full ED to calculate the specific heat in the $t$-$J$ model with $N$ sites, incorporating all-to-all random interactions and hoppings,
\begin{equation}
\mathcal{H} = \frac{1}{\sqrt{N}}\sum_{i\neq j=0}^{N-1} t_{i j} P c^\dagger_{i\alpha} c_{j\alpha} + \frac{1}{\sqrt{N}} \sum_{i< j=0}^{N-1} J_{ij} \boldsymbol{S}_i \cdot \boldsymbol{S}_j,
\end{equation}
where $P$ is the projection on non-doubly occupied sites, $\boldsymbol{S}_i=\frac{1}{2}c^\dagger_{i\alpha} \boldsymbol{\sigma}_{\alpha \beta}c_{j\beta}$ is the spin operator on site $i$. Both the hoppings $t_{ij}=t^\ast_{ji}$ and the exchange interaction $J_{ij}$ are independent random numbers with zero mean and variance $\bar{t^2}$ and $\bar{J^2}$, respectively. This type of system exhibits a transition from a spin glass to a disordered Fermi liquid phase with increasing doping, as shown in Ref.~\cite{Shackleton2021}. 

\paragraph{Specific heat, triangular $t$-$J$ model}
We perform full ED to calculate the specific heat \(C\) of the triangular lattice t-J model with the Hamiltonian
\begin{equation}
\mathcal{H} = -t \sum_{\langle i, j\rangle, \sigma} P c^\dagger_{i\alpha} c_{j\alpha} + J \sum_{\langle i,j \rangle} \left(\boldsymbol{S}_i \cdot \boldsymbol{S}_j -\frac{1}{4}n_{i}n_{j} \right)
\end{equation}

\noindent with $J/t=0.4$. We consider an $11$-site triangular lattice (arranged in a 4-4-3 geometry) with one hole in the system. The number of electrons is thus $N-1$ (with equal numbers of spin-up and spin-down electrons).

\paragraph{Level statistics in spin $S=1/2$ chains}
This example demonstrates the differing level statistics of integrable and non-integrable models using the example of the $S=1/2$ XXZ chain with and without next-nearest neighbor interactions. The term "level statistic" refers to the probability distribution $P(s)$ where the variable $s$ is the difference between adjacent energy levels $0 \leq E_{i+1}-E_i$. That is, for a given quantum system, $P(s)$ describes the likelihood that the next excited state above a randomly chosen energy level $E$ has energy $E + s$. For integrable systems, the level statistic (after a renormalization by the mean energy) is a Poissonian
\begin{equation}
P_{\mathrm{Poiss}}(s) = \exp(-s)
\end{equation}
whereas the so-called Wigner-Dyson distribution emerges for non-integrable systems
\begin{equation}
P_{\mathrm{WD}}(s) = \frac{\pi s}{2} \exp(- \pi s^2/4).
\end{equation}

\paragraph{Many-body localization}
We consider a transverse field Ising model with random interaction and field strengths,
\begin{equation}
    \mathcal{H} = \sum_i J_i \sigma^z_i \sigma^z_{i+1} - \sum_i (h_i\sigma_i^z + \gamma_i\sigma_i^x),
\end{equation}
where we take $J_i\in [0.8,1.2]$, $h_i\in [-W,W]$, $\gamma_i=1$, and $W$ is the disorder strength. This model is known to display many-body localization for large enough $W$, i.e, it hosts an extensive set of emergent (local) conserved quantities, often called $l$-bits. To study this, we compute the inverse participation ratio, $\mathrm{IPR}=\sum_i |\psi_i|^4$, most often used to study Anderson localization, for some of the low-lying eigenstates. Additionally, we also study the level spacing ratio, $r=\text{min}(\delta_n,\delta_{n+1})/\text{max}(\delta_n,\delta_{n+1})$, where $\delta_n=E_{n+1}-E_n$.

\paragraph{Lindblad Super-operator spectrum}
We obtain the spectrum of the Lindblad super-operator governing the time-evolution of an $S=1/2$ XXZ chain with bulk dephasing~\cite{PhysRevE.92.042143},
\begin{equation}
    \mathcal{H} = J \sum_i \left( S^x_iS^x_{i+1} + S^y_{i+1}S^y_{i} + \Delta S^z_i S^z_{i+1} \right),\; L_i =\sqrt{\gamma} S^z_i,
\end{equation}
where $L_i$ is the quantum jump operator associated with a Markovian bath acting on site $i$ and $\gamma$ the strength of the dissipation. The spectrum is obtained by constructing the vectorized Lindblad operator,
\begin{equation}
\begin{aligned}
\mathcal{L}
= \sum_{n=0}^{L-1} \Bigl[
& -iJ \Bigl(
    S_{n}^{+} S_{n+1}^{-}
  + S_{n+1}^{+} S_{n}^{-}
  + \Delta\, S_{n}^{z} S_{n+1}^{z}
\Bigr) \\
& + iJ \Bigl(
    \tilde{S}_{n}^{+} \tilde{S}_{n+1}^{-}
  + \tilde{S}_{n+1}^{+} \tilde{S}_{n}^{-}
  + \Delta\, \tilde{S}_{n}^{z} \tilde{S}_{n+1}^{z}
\Bigr)
\Bigr] \\
& + \gamma \sum_{n=0}^{L-1}
\left(
  S_{n}^{z} \tilde{S}_{n}^{z}
  - \frac{i}{4}
\right),
\end{aligned}
\end{equation}
where $\tilde{S}$ are auxiliary spin degrees of freedom.

The Lindblad spectrum reveals the system’s excitations and decay timescales. A key quantity is the dissipative gap, defined as the absolute value of the real part of the eigenvalue closest to zero. This gap sets the slowest relaxation time, and its dependence on the model parameters plays a central role in the study of dissipative phase transitions~\cite{PhysRevA.98.042118,soares2025dissipativephasetransitioninteracting}.

\subsection{Tower-of-states}

\paragraph{Symmetry-breaking $\alpha \text{XX}$ chain}
We perform a tower-of-states analysis of the $\alpha \text{XX}$ model~\cite{PhysRevA.105.022625}. This $S=1/2$ model consists of an array of $N$ sites with long-range $\text{XY}$  couplings that decay as a power law with exponent $\alpha$,
\begin{equation}
    \mathcal{H} = -\dfrac{J}{2} \sum_{i<j} \dfrac{S^+_iS^-_j+S^-_iS^+_j}{|i - j|^\alpha}.
\end{equation}
The TOS analysis provides strong evidence for the existence of spontaneous symmetry breaking (SSB) in the thermodynamic limit. Otherwise, the ground state of a finite system is always completely symmetric.

\paragraph{Heisenberg chain momentum-resolved spectrum}
This example demonstrates how to set up a TOS analysis using translation symmetry for the example of a spin $S=1/2$ Heisenberg chain from Eq.~\eqref{eq:heisenberg_model}, with $J>0$ taken as antiferromagnetic.

\paragraph{Square lattice Heisenberg model}
We perform a TOS analysis of the Heisenberg model on a square lattice with periodic boundary conditions. This model consists of spin $S=1/2$ sites at the vertices of a square lattice with nearest-neighbor antiferromagnetic Heisenberg couplings. The TOS analysis provides strong evidence for SSB in the thermodynamic limit, as the ground state of a finite system is completely symmetric. 
By plotting the spectrum obtained using XDiag vs. $S_\text{tot}(S_\text{tot}+1)$, defined as the eigenvalues of the total spin operator,
\begin{equation}
    \hat{\mathbf{S}}_{\mathrm{tot}} = \sum_{i} \hat{\mathbf{S}}_i,
\qquad
\hat{\mathbf{S}}_{\mathrm{tot}}^{2} = \hat{\mathbf{S}}_{\mathrm{tot}}\cdot\hat{\mathbf{S}}_{\mathrm{tot}},
\end{equation}
we show that the low-energy spectrum of this model scales linearly with $S_\text{tot}(S_\text{tot}+1)$, a telltale sign of antiferromagnetic order.

\paragraph{Triangular lattice $J_1$-$J_2$ model}
We perform a TOS analysis of the $J_1$-$J_2$ spin $S=1/2$ Heisenberg model on the triangular lattice~\cite{Wietek2016b} with nearest and next-nearest neighbor Heisenberg interactions,
\begin{equation}
    \mathcal{H} = J_1 \sum_{\langle i, j \rangle} \boldsymbol{S}_i \cdot \boldsymbol{S}_j + J_2 \sum_{\langle\langle i, j \rangle\rangle} \boldsymbol{S}_i \cdot \boldsymbol{S}_j.
\end{equation}
Depending on the ratio $J_2/J_1$, this model can stabilize several magnetic orders on the triangular lattice.

\paragraph{Extended kagome lattice antiferromagnet}
We consider the kagome lattice equipped with the spin $S=1/2$ Hamiltonian
\begin{equation}
\mathcal{H} = J_1\sum_{\langle i,j\rangle} \boldsymbol{S}_i \cdot \boldsymbol{S}_j+J_2\sum_{\langle \langle i,j\rangle\rangle} \boldsymbol{S}_i \cdot \boldsymbol{S}_j+J_3\sum_{\langle \langle\langle i,j\rangle\rangle\rangle_h} \boldsymbol{S}_i \cdot \boldsymbol{S}_j,
\end{equation}
where $\langle \dots \rangle$ and $\langle\langle \dots \rangle\rangle$ denote sums over nearest and next-nearest neighbor sites, and $\langle \langle\langle\dots\rangle\rangle\rangle_h$ denotes a sum over third-nearest-neighbor sites~\cite{Wietek2015,Wietek2020}. This model hosts a plethora of interesting phases for different parameter regimes. We focus on the parameter values $J_2/J_1 =-1.0$ and $J_3/J_1=-2.0$, which were predicted~\cite{Wietek2020} to realize a spin-nematic order with quadrupolar character. By plotting the low-energy spectrum, we show that odd-$S_\text{tot}$ sectors are not present in the low-energy TOS, indicating a quadrupolar spin-nematic phase.

\paragraph{Square lattice attractive Hubbard model}
We perform a TOS analysis on the attractive Hubbard model on a square lattice. This model is known to display a superconducting ground state, which becomes apparent by a degeneracy in even-number particle sectors in the spectrum, signaling $U(1)$ symmetry breaking.

\subsection{Dynamical spectral functions}

\paragraph{Dynamical spin structure factor}
This example uses the Lanczos algorithm to calculate the dynamical spin spectral function of the spin $S=1/2$ Heisenberg chain. The structure factor is defined as
\begin{equation}
    S^{zz}({\bf k},\omega) = \int dt e^{-i\omega t}\langle S^z_{\bf k}(t)S^z_{\bf -k}\rangle,
\end{equation}
where $S^z_{\bf k}=\frac{1}{\sqrt{N}}\sum_i e^{-i\bf{k\cdot r}_i}S^z_i$.

\paragraph{Hubbard model Green's function}
This example uses the Lanczos algorithm to calculate the Green's function of the Hubbard model. The Green's function is given by
\begin{equation}
    G({\bf k}, \omega)=-i\int dt e^{-i\omega t}\langle \lbrace c_{\bf k}(t),c^{\dagger}_{\bf k}\rbrace\rangle.
\end{equation}
To achieve a mesh of momentum space, we use the generated momenta inside the Wigner-Seitz cell allowed by the finite cluster or twisted boundary conditions~\cite{poilblanc1991}.

\paragraph{Optical conductivity $t$-$J$ model}
We use the Lanczos algorithm to calculate the conductivity of the planar $t$-$J$ model with one hole. The ground state conductivity is given by the current autocorrelator
\begin{equation}
    \sigma(\omega)=-i\int dt e^{-i\omega t}\langle  J(t)J\rangle.
\end{equation}
The plotting script includes a simple benchmark exploring the role of the number of Lanczos iterations on the result.

\subsection{Time evolution}

\paragraph{Domain wall dynamics}
This example demonstrates the time evolution of a domain wall in the ferromagnetic spin $S=1/2$ XXZ chain. Note that here we use open instead of periodic boundary conditions, allowing to have a single domain wall in the system. We choose as the initial state the following eigenstate of the associated Ising chain
\begin{equation}
|\psi_0\rangle= |\uparrow\ldots\uparrow\downarrow\ldots\downarrow\rangle,
\end{equation}
featuring a domain wall at the center, which melts over time.

\paragraph{Bubble nucleation in the mixed-field Ising chain}

We study the dynamics of the one-dimensional Ising chain under two external fields,
\begin{equation}
\mathcal{H} =
  - \sum_{n=1}^{N-1} S^z_n S^z_{n+1}
  - \frac{h_x}{2} \sum_{n=1}^{N} S^x_n
  - \frac{h_z(t)}{2} \sum_{n=1}^{N} S^z_n,
\end{equation}
where $h_x$ introduces quantum fluctuations and $h_z(t) = h_z^{\text{in}} + t/\tau_Q$ is the time-dependent longitudinal field. This system exhibits a first-order quantum phase transition (FOQPT) between two symmetry-broken ferromagnetic phases with opposite longitudinal magnetization. When the longitudinal field is slowly ramped from negative to positive values across the FOQPT, the system undergoes quantized bubble nucleation events~\cite{Sinha2021}: sharp nonadiabatic transitions in the dynamics where local domains (or “bubbles”) of the new vacuum appear within the metastable phase.

\subsection{Thermodynamics}

\paragraph{Shastry-Sutherland model}
We use an ensemble of thermal pure quantum states~\cite{Sugiura2013} (or finite-temperature Lanczos method~\cite{Prelovsek1994}) to compute the total energy and specific heat as functions of temperature for the Shastry–Sutherland model throughout the dimer phase, as performed in Ref.~\cite{Wietek2019}

\paragraph{Square $J_1$-$J_2$ Heisenberg model}
In this example, we compute various thermodynamic quantities of the square lattice $J_1$-$J_2$ model, including susceptibility, entropy, and the Wilson ratio. We again use thermal pure quantum states~\cite{Sugiura2013} or the finite-temperature Lanczos method~\cite{Prelovsek1994} and explore the phase diagram of the model. The Wilson ratio is shown to signal the onset of quantum paramagnetic behavior.

\section{Implementation}
\label{sec:implementation}

We present the core design principles of the XDiag library in the following sections. The full source code of XDiag is available on GitHub~\cite{github}.

\subsection{Basis objects}
The core functionality of XDiag is provided by the block objects of types like \lstinline{Spinhalf}, \lstinline{tJ}, and \lstinline{Electron}. While these carry information about quantum numbers, they carry a reference to a so-called \textit{Basis} object. The basis objects implement how exactly we encode the block of the Hilbert space. This includes which integer type is used for encoding product states, as well as how the product states are ordered. Basis objects can store significant amounts of data, including the product state configurations, norms of symmetrized states, and also lookup tables for efficient use of symmetries. Basis objects are not directly accessible to users and are created dynamically at runtime. Because block objects only carry a reference (implemented as a \lstinline{std::shared_ptr} shared pointer to a \lstinline{std::variant} of bases), they can be efficiently copied without much overhead while copying bases is avoided internally. 

When constructing a block, an optional argument \lstinline{backend} can be handed to the constructor which selects the basis type used. By default, this argument is set to \lstinline{"auto"}, and XDiag decides automatically which backend is chosen. Typical arguments are \lstinline{"32bit"} and \lstinline{"64bit"}; this allows one to choose whether 32-bit integers or 64-bit integers are used to represent basis states. Moreover, the \lstinline{Spinhalf} block features the \lstinline{"2sublattice"}, \lstinline{"3sublattice"}, \lstinline{"4sublattice"}, and \lstinline{"5sublattice"} backends which implement the sublattice coding algorithms proposed in Ref.~\cite{Wietek2018}.

Internally, the type of the block then determines which implementations are used dynamically at runtime. We entirely avoid exposing any template arguments to the user interface. While static polymorphism using templates is used to implement different low-level routines, on a higher level, different implementations are chosen at runtime without incurring any computational overhead. We demonstrate this design principle using the \lstinline{apply} function, one of the core functionalities of XDiag. While at the user interface level, its signature is given by,

\begin{lstlisting}[language=C++]
State apply(Op const &op, State const &v);
\end{lstlisting}
calling this routine will be channeled down to a low-level call to

\begin{lstlisting}[language=C++]
template <typename mat_t, typename block_t>
void apply(OpSum const &ops, 
           block_t const &block_in, mat_t const &mat_in,
           block_t const &block_out, mat_t &mat_out);
\end{lstlisting}
which then decides the type of implementation used. Here, we template over the matrix type of the coefficient vector, which can be the Armadillo types \lstinline{arma::vec}, \lstinline{arma::cx_vec}, \lstinline{arma::mat}, \lstinline{arma::cx_mat}, and also the type of the block. Further down the call stack, we then decide which implementation to choose based on the type of the basis object referenced to by the block object. A typical function called is the \lstinline{apply_terms} function, which is called both when applying operators to states but also when creating matrices of operators using the \lstinline{matrix} function.

\begin{lstlisting}[language=C++]
template <typename coeff_t, bool symmetric, class basis_t, 
          class fill_f>
void apply_terms(OpSum const &ops, basis_t const &basis_in,
                 basis_t const &basis_out, fill_f fill);
\end{lstlisting}
Here, we template over the coefficient type \lstinline{coeff_t}, which can be either \lstinline{double} or \lstinline{std::complex<double>}, a boolean deciding if our computation uses symmetries, the type of the basis, and also the fill function as a functor. The fill function either fills in a matrix with the computed coefficients or fills an output vector from an input vector. 

\subsection{Indexing}
Key algorithmic problems in ED are enumerating basis configurations and retrieving the index of a basis configuration. These two fundamental problems have, in the past, been addressed in several works, where we have now implemented several ideas in XDiag. Lin tables~\cite{Lin1990} offer an efficient way of retrieving the index of a spin configuration with a given particle number and are employed in most block types whenever no permutation symmetries are used. 

Whenever permutation symmetries are employed, we use fast lookup tables to store the action of a symmetry on a configuration and also to store the representative of a given spin configuration, see e.g.~\cite{Weisse2008,WietekThesis} for detailed definitions of symmetry-adapted wave functions. While for \lstinline{tJ} and \lstinline{Electron} type blocks, these lookup tables can be stored with reasonable memory overhead, they can become prohibitively large in the case of \lstinline{Spinhalf} blocks. For this case, we provide the sublattice coding algorithms described in Ref.~\cite{Wietek2018} to allow for fast and memory-efficient implementations of symmetry-adapted bases. 

At several points, a search of a product state within the list of basis representations is necessary. This search is divided into a lookup on prefix states by using a hash table and a binary search on postfix states. This way, we combine the speed of hash tables with the memory efficiency of a binary search. While testing several implementations of hash tables, we found the \lstinline{flat_hash_map}~\cite{flat_hash_map} to be both favorable in speed and memory efficiency as compared to other implementations. 

In implementations of the \lstinline{tJ} and \lstinline{Electron} blocks, we make use of the product structure of these Hilbert spaces, separating $\uparrow$-spin configurations from $\downarrow$-spin configurations. Conventional spin-conserving hopping terms, for example, can be applied to a single spin species without modifying the other. This is used to speed up Hilbert space searches. Moreover, whenever using permutation symmetries, we take this structure into account, which can lead to improved searches of representatives and their corresponding symmetries, as only subgroups of the full permutation group can be involved. A particularly interesting optimization option is used in the \lstinline{tJ} block, where we employ the novel \lstinline{pext} and \lstinline{pdep} machine instructions of the x86 bit manipulation instruction set~\cite{bmi} to interleave $\uparrow$-and $\downarrow$-spin configurations.

\subsection{Parallelization}
XDiag offers two different parallelizations. A shared memory parallelization is provided in the standard \lstinline{xdiag} library using \lstinline{OpenMP}~\cite{openmp}. We parallelize individual matrix-vector applications and matrix generations and demonstrate in section~\ref{sec:benchmarkshared} that our implementation scales close to optimal up to $N=64$ threads. Parallelism for linear algebra routines such as dot products or full matrix diagonalizations is not part of the XDiag library, but can be obtained by linking to threaded linear algebra backends. By default, when building the standard library, we look for an existing installation of the Intel MKL library~\cite{mkl}, which provides thread-level parallelism on all linear algebra routines. Another common choice of a threaded linear algebra backend is OpenBLAS, which is the standard choice in Julia. The shared memory parallelization is kept rather simple, and we parallelize iterating over basis states. We take particular care whenever atomic operations are necessary and when they can be avoided, depending on the details of the computation. Besides a few instances of atomic operations, no synchronization between threads is found to be necessary.

A distributed memory parallelization of elementary operations using the message passing interface (MPI)~\cite{mpi} is provided in the \lstinline{xdiag_distributed} library. The implementation is more involved and requires a strategy of distributing computational basis states among the MPI processes. We adopt the strategy proposed in Ref.~\cite{Wietek2018}, where basis states are distributed according to a hash function among the processes. This strategy is used throughout our different blocks, and we find that this way, a satisfying strong scaling up to several thousands of cores can be achieved, cf. section~\ref{sec:benchmarkdistributed}. Moreover, the choice of configurations to be distributed is adapted to the Hilbert space structure such that large parts of the computation can be performed locally without any need for communication between the processes. In the case of the \lstinline{ElectronDistributed} block, we use a particular ordering, in which all $\downarrow$-spin configurations belonging to a single $\uparrow$-spin configuration are located on one MPI process. With this memory layout, we can then apply, e.g., a hopping term on the $\downarrow$-spins locally without communication. Similarly, we define the reverse order, where all $\uparrow$-spin configurations belonging to a single $\downarrow$-spin configuration are stored on one process that is then used to apply hopping on $\uparrow$-spins. In total, for a Hubbard-type model, this only requires communication for two transpose operations to change to reverse order and back. Similar ideas are implemented for the \lstinline{SpinhalfDistributed} and \lstinline{tJDistributed} blocks. In the former case, we distinguish between prefix spin configurations and postfix spin configurations. If an operator is only acting on the postfix configurations, no communication is required, and analogously, a transpose operation is performed to act on prefix configurations. 

\subsection{Julia wrapper}
While all core functionality is implemented in the C++ library, we provide a convenient Julia wrapper to make XDiag functionalities readily available without the need to compile the library and application codes. We used the \lstinline{CxxWrap.jl} package~\cite{cxxwrap} to write the wrapper code on the C++ side. The XDiag library, including the wrapper code, is then compiled into shared libraries using the \lstinline{BinaryBuilder.jl}~\cite{binarybuilder}. The binaries for various platforms and corresponding \lstinline{XDiag_jll.jl} package are generated via the \lstinline{Yggdrasil} framework~\cite{yggdrasil} by performing a pull request to its GitHub repository, where the build recipe for XDiag can be found. Finally, a Julia package \lstinline{XDiag.jl} has been registered in the Julia registry, and thus installing XDiag in Julia is as simple as entering package mode and adding XDiag. 

\begin{lstlisting}[language=Julia]
pkg> add XDiag
\end{lstlisting}

\noindent Having precompiled binaries for the core functionality leads to the positive side effect that precompilation times are minimal, as only the thin wrapper layer needs to be precompiled in Julia. 

\section{Benchmarks}
\label{sec:benchmarks}

We benchmark the computation time of the key operation behind all main features implemented in XDiag: the execution of a single Lanczos iteration, consisting of a matrix-vector multiplication and a few vector operations, and the generation of the Hamiltonian matrix as a sparse matrix in CSR format. We investigate the performance of all possible Hilbert spaces supported by XDiag, namely the \lstinline{Spinhalf}, \lstinline{Electron}, and \lstinline{tJ} blocks and, thus, consider three distinct models to perform the benchmarks: 
\begin{enumerate}
\item The $S=1/2$ one-dimensional Heisenberg chain with nearest-neighbor interactions, 
\begin{equation}
H= J \sum_{\langle i,j\rangle} \mathbf{S}_{i} \cdot \mathbf{S}_{j},
\label{eq:heisenberg_model2}
\end{equation}
as previously introduced in Eq.~\ref{eq:heisenberg_model}.
\item The one-dimensional Hubbard model,
\begin{equation}
H= -t\sum_{\langle i,j\rangle\; \sigma}\left(c^\dagger_{i,\sigma}c_{j,\sigma} + \text{h.c.} \right) + U\sum_{i=0}^{N} n_{i,\uparrow} n_{i,\downarrow},
\label{eq:hubbard}
\end{equation}
where $t$ is the hopping strength between first-neighboring sites, $c^\dagger_{i,\sigma},  c_{i,\sigma} $ the creation and annihilation operator which creates (resp. destroys) a fermion on site $i$ with spin $\sigma=\uparrow,\downarrow$ and $U$ the strength of the on-site density-density interaction.
\item The t-J model of $S=1/2$ electrons, 
\begin{equation}
H= -t\sum_{\langle i,j\rangle\; \sigma} P\left(c^\dagger_{i,\sigma}c_{j,\sigma} + \text{h.c.} \right)P + J\sum_{\langle i,j\rangle} \left( \boldsymbol{S}_i \cdot \boldsymbol{S}_{j} - \dfrac{n_{i} n_{i+1}}{4} \right),
\label{eq:tj}
\end{equation}
where $P$ projects out doubly occupied sites and $n_i= n_{i,\uparrow} + n_{i,\downarrow}$.
\end{enumerate}

XDiag supports both multithreading using OpenMP and multiprocessing using MPI. As such, we provide benchmarks for the three models using both. We emphasize that in the current version of XDiag, lattice symmetries are not supported when using MPI parallelization. Beyond parallel scalability, we also assess the efficiency against established libraries: we compare the serial performance of XDiag's Lanczos routine with that of the ALPS library~\cite{Albuquerque2007}, and benchmark the sparse matrix generation against the QuSpin library~\cite{Weinberg2017,Weinberg2019}.

In all the following sections, the computation time always corresponds to the wall time and not to the CPU time.
\begin{figure}
    \centering
    \makebox[\textwidth]{\includegraphics{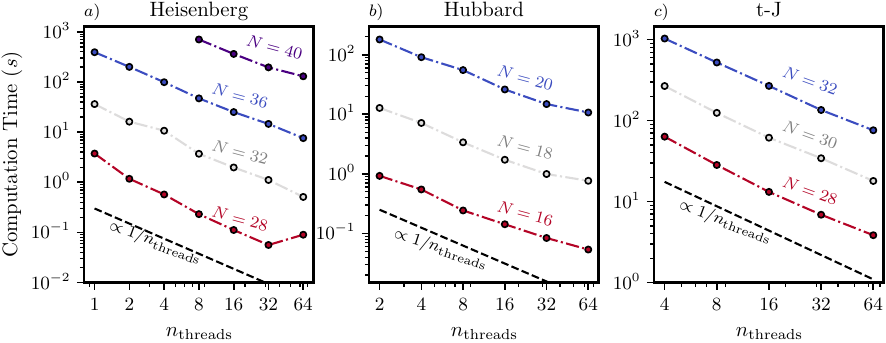}}
    \caption{Scaling of the computation time for a single Lanczos iteration as a function of the total number of threads used. The calculations were done for the Heisenberg $a)$, Hubbard $b)$ and t-J $c)$ models on a linear chain using different total numbers of spins/sites. The dashed black line in each panel represents the slope corresponding to ideal linear scaling. In the calculation, we used both the U(1) and lattice symmetries.}
    \label{fig:bench_1dheisen}
\end{figure}

\subsection{Shared memory parallelization}
\label{sec:benchmarkshared}

We show how the computation time of a single Lanczos step depends on the total number of OpenMP threads used for the shared memory parallelization as available. For the \lstinline{Spinhalf} block we consider system sizes of $N=28$, $N=32$, $N=36$ and $N=40$ at zero magnetization, for the \lstinline{Electron} block we consider system sizes of $N=16$, $N=18$, and $N=20$ at half-filling, and for the \lstinline{tJ} block we consider system sizes of $N=28$, $N=30$, and $N=32$ in the presence of two holes. In addition to the U(1) symmetry, associated with the conservation of total magnetization in the Heisenberg model, or the conservation of total particle number in the Hubbard and t-J models, we also exploit the lattice translation symmetries to perform the calculations (see Section~\ref{subsec:symmetries}). Here, the Hamiltonian matrix elements are computed on-the-fly to maximize the memory available for storing the Lanczos vectors. The benchmarks presented were performed on a single node equipped with an Intel Xeon(R) Platinum $8360\text{Y}$ (Ice Lake) 2.4 GHz processor.

In Fig.~\ref{fig:bench_1dheisen}, we present the scaling of the computational time of a single Lanczos step with the total number of threads used in the calculation. As shown in the three panels, the computation time practically follows a linear scaling with the number of OpenMP threads, decreasing with the inverse of the total number of threads. In table~\ref{tab:threads_computational_time}, we present the computational time for the calculation using a given type of Hilbert space block and the symmetries used with $n_{\textrm{threads}}=64$ threads. We conclude that the multithreading parallelization of XDiag exhibits close to linear scaling up to $n_{\textrm{threads}}=64$ threads.


\begin{table}
  \centering
  \begin{tabular}{ccccc}
    \toprule
     {Block} & {$N$} & {Symmetries} & {Block Size} & {Time (secs.)} \\
    \midrule
    \lstinline[]$Spinhalf$ &40 & U(1) and Lattice &  $1.7 \cdot 10^{9}$  & $131 \pm 1 $ \\
 \lstinline[]$Spinhalf$ &36 & U(1) and Lattice &  $1.3 \cdot 10^{8}$  & $7.60 \pm 0.2 $ \\
    \lstinline[]$Spinhalf$ &32 & U(1) and Lattice &  $9.3 \cdot 10^{6}$  & $\left(5.1 \pm 0.1 \right)\cdot 10^{-1}$ \\
    \lstinline[]$Spinhalf$ &28 & U(1) and Lattice &  $7.2\cdot 10^{5}$  & $\left(9 \pm 2 \right)\cdot 10^{-2}$ \\
    \lstinline[]$Electron$ &22 & U(1) and Lattice &  $2.2 \cdot 10^{10}$  & $138 \pm 5 $ \\
    \lstinline[]$Electron$ &20 & U(1) and Lattice &  $1.0 \cdot 10^{9}$  & $10.9 \pm 0.1 $ \\
    \lstinline[]$Electron$ &18 & U(1) and Lattice &  $1.3 \cdot 10^{8}$  & $7.64 \pm 0.01 $ \\
    \lstinline[]$Electron$ &16 & U(1) and Lattice &  $1.0 \cdot 10^{7}$  & $5.37 \pm 0.01 $ \\
    \lstinline[]$tJ$ &32 & U(1) and Lattice  & $2.0 \cdot 10^{9}$ & $75.9 \pm 0.1 $ \\
    \lstinline[]$tJ$ &30 & U(1) and Lattice  & $5.8 \cdot 10^{8}$ & $17.95 \pm 0.05$ \\
    \lstinline[]$tJ$ &28 & U(1) and Lattice  & $1.4 \cdot 10^{8}$ & $3.84 \pm 0.02 $ \\
    \bottomrule
  \end{tabular}
    \caption{Computation times for a single Lanczos interaction for each block type using certain symmetries. In all results, we have used $n_\text{threads}=64$.}
  \label{tab:threads_computational_time}
\end{table}

\subsection{Distributed memory parallelization}
\label{sec:benchmarkdistributed}

\begin{figure}
    \centering
    \makebox[\textwidth]{\includegraphics{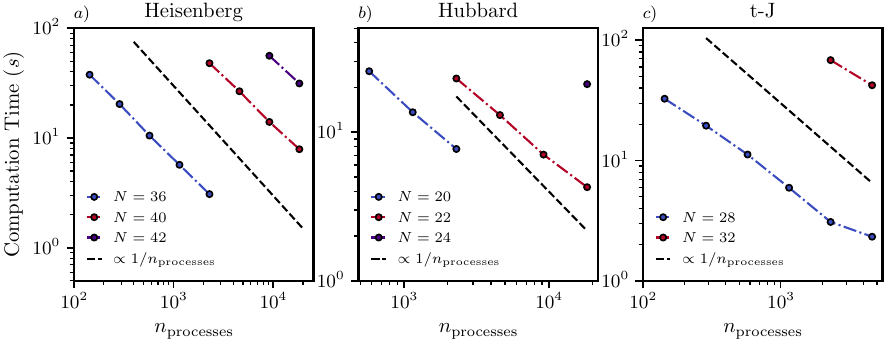}}
    \caption{Scaling of the computation time for a single Lanczos iteration as a function of the total number of MPI processes used. The calculations were done for the Heisenberg $a)$, Hubbard $b)$, and t-J $c)$ models using different total numbers of spins/sites. The dashed black line in each panel represents the slope corresponding to ideal linear scaling. In $a)$, the calculation was performed in the zero magnetization sector, in $b)$ it was done at half-filling for $N =20$, at $3/11$ for $N=22$ and $1/3$ for $N=24$.}
    \label{fig:bench_mpi}
\end{figure}

XDiag also supports distributed-memory parallelization via MPI for the \lstinline{SpinhalfDistributed}, \lstinline{ElectronDistributed}, and \lstinline{tJDistributed} blocks. We benchmark the computation time of a single Lanczos Iteration as a function of the number of MPI processes used. The calculations were performed on the Max-Planck-Gesellschaft Raven supercomputer using nodes also equipped with Intel Xeon(R) Platinum $8360\text{Y}$ (Ice Lake) 2.4 GHz processors, each featuring 72 cores. These nodes are interconnected with a 100 $\text{Gb}/\text{s}$ HDR100 network. In these benchmarks, lattice symmetries are not used, only U(1) number conservation symmetries. We consider the case of $N=36$ and $N=40$ for the Heisenberg model at zero magnetization. For the Hubbard model, we consider $N=20$ at half-filling, and $N=22$ and $N=24$ with $n_{\uparrow}=n_{\downarrow}=8$ electrons. The $t$-$J$ model is studied on $N=28$ and $N=32$ sites with $2$ holes at zero magnetization. We observe that for all systems considered, the computation time scales almost inversely proportional to the total number of MPI processes, as shown in Fig,~\ref{fig:bench_mpi}. Moreover, we provide, in the table~\ref{tab:threads_computational_time_mpi}, the actual computational time required for each Lanczos iteration across the different models considered. We conclude that the MPI parallelization exhibits almost linear strong scaling up to several thousand processes.

 \begin{table}
  \centering
  \begin{tabular}{cccccc}
    \toprule
     {Block} & {$N$} & {Symmetries} & {Block Size} & { MPI processes} & {Time (secs.)} \\
    \midrule
     \lstinline[]$Spinhalf$ &42 & U(1) &  $5.4 \cdot 10^{11}$  & 18432 &$31.4\pm 0.5 $ \\
 \lstinline[]$Spinhalf$ &40 & U(1) &  $1.4 \cdot 10^{11}$  & 18432 &$7.9 \pm 0.1 $ \\

  \lstinline[]$Spinhalf$ &36 & U(1) &  $9.0 \cdot 10^{9}$  & 2304 &$3.1 \pm 0.2 $ \\
    \lstinline[]$Electron$ &24 & U(1) &  $5.4 \cdot 10^{11}$ & 18432& $21.0 \pm 0.2 $ \\
     \lstinline[]$Electron$ &22 & U(1) &  $1.0 \cdot 10^{11}$ &  18432 & $4.3 \pm 0.1 $ \\
    \lstinline[]$Electron$ &20 & U(1) &  $3.4 \cdot 10^{10}$ &  2304 & $7.7 \pm 0.5 $ \\
    \lstinline[]$tJ$ &32 & U(1)  & $7.6\cdot 10^{10}$ & 4608&$42.2 \pm 0.5 $ \\
    \lstinline[]$tJ$ &28 & U(1)  & $3.9\cdot 10^{9}$ & 4608&$2.3 \pm 0.4 $ \\
    \bottomrule
  \end{tabular}
    \caption{Computation times for a single Lanczos interaction for each block type using a different number of MPI processors.}
  \label{tab:threads_computational_time_mpi}
\end{table}

\subsection{Sparse Matrix Generation}
\label{sec:comparison_other_libraries}
\begin{figure}
    \centering
    \makebox[\textwidth]{\includegraphics{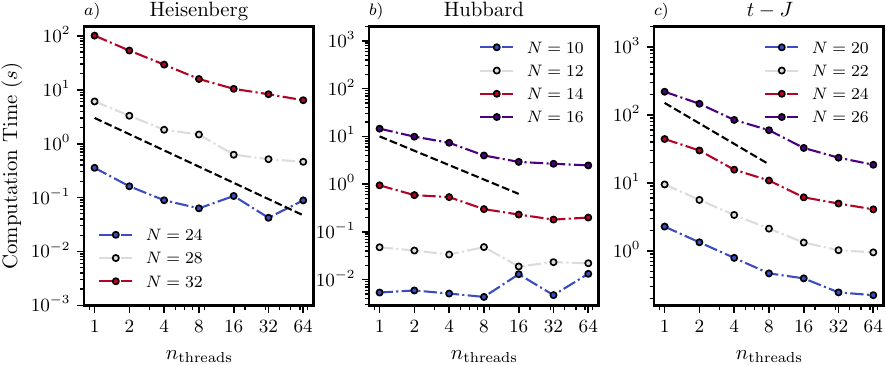}}
    \caption{Computational time for the generation of the Hamiltonian matrix in the sparse CSR matrix format as a function of the total number of threads used. The calculations were done for the Heisenberg $a)$, Hubbard $b)$, and t-J $c)$ models using different total numbers of spins/sites. The dashed black line in each panel represents the slope corresponding to ideal linear scaling.}
    \label{fig:benchmark_sparce}
\end{figure}
XDiag supports the creation of several common sparse matrix formats. In particular, the CSR format can be used directly by the library's iterative solvers. In this section, we analyze the computation time required to generate the CSR matrix and evaluate its scalability with respect to the number of OpenMP threads. We consider the Hamiltonian for three distinct models with the following system sizes: (i) \lstinline{Spinhalf} with $N=24$, $N=28$, and $N=32$; (ii) \lstinline{Electron} with $N=10$, $N=12$, $N=14$, and $N=16$; and (iii) \lstinline{tJ} with $N=20$, $N=22$, $N=24$, and $N=26$. In all calculations, we exploit again both lattice translation symmetries and the relevant U(1) symmetry. Specifically, we consider the $S^z=0$ sector for the Heisenberg model, half-filling block for the Hubbard model, and the two-hole sector with $S^z=0$ for the t-J model. These benchmarks were performed on a single node equipped with an Intel Xeon Platinum 8360Y (Ice Lake) 2.4~GHz processor.

\begin{table}
  \centering
  \begin{tabular}{ccccc}
    \toprule
     {Block} & {$N$} & {Symmetries} & {Block Size} & {Time (secs.)} \\
    \midrule
    \lstinline[]$Spinhalf$ &32 & U(1) and Lattice &  $9.4 \cdot 10^{6}$  & $(6.40 \pm 4)\cdot10^3 $ \\
 \lstinline[]$Spinhalf$ &28& U(1) and Lattice &  $7.2 \cdot 10^{5}$  & $(4.6 \pm 2)\cdot10^2 $ \\
    \lstinline[]$Spinhalf$ &24 & U(1) and Lattice &  $5.6 \cdot 10^{4}$  & $(8.8 \pm 2)\cdot 10 $ \\
    \lstinline[]$Electron$ &16 & U(1) and Lattice &  $1.0 \cdot 10^{7}$  & $(2.47 \pm 0.02)\cdot 10^{3} $ \\
    \lstinline[]$Electron$ &14 & U(1) and Lattice &  $8.4 \cdot 10^{5}$  & $(1.99 \pm 0.05)\cdot 10^2 $ \\
    \lstinline[]$Electron$ &12 & U(1) and Lattice &  $7.1 \cdot 10^{4}$  & $(2.2 \pm 0.1)\cdot 10 $ \\
    \lstinline[]$Electron$ &10 & U(1) and Lattice &  $6.4 \cdot 10^{3}$  & $(1.32 \pm 0.07)\cdot 10 $ \\
    \lstinline[]$tJ$ &26 & U(1) and Lattice  & $3.4 \cdot 10^{7}$ & $( 1.845 \pm 0.008) \cdot 10^4$ \\
    \lstinline[]$tJ$ &24 & U(1) and Lattice  & $8.1 \cdot 10^{6}$ & $( 4.070 \pm 0.005) \cdot 10^3$ \\
    \lstinline[]$tJ$ &22 & U(1) and Lattice  & $1.9 \cdot 10^{6}$ & $(9.5 \pm 0.1)\cdot 10^{2} $ \\
    \lstinline[]$tJ$ &20& U(1) and Lattice  & $4.6 \cdot 10^{5}$ & $(2.22\pm0.07) \cdot 10$ \\
    \bottomrule
  \end{tabular}
    \caption{Computation times for the generation of the Hamiltonian matrix in the CSR format for each block type when $n_\text{threads}=64$.}
  \label{tab:threads_computational_time_sparce_matrix}
\end{table}

Fig.~\ref{fig:benchmark_sparce} illustrates the scaling of computational time as a function of the number of OpenMP threads. For larger system sizes, we observe a significant decrease in the computational time as the number of threads increases. Notably, for the Heisenberg model $a)$, the computational time scales almost linearly with the inverse of the thread number up to $n_{\rm threads}=8$. In contrast, the Hubbard $b)$ and $t-J$ $c)$ models exhibit a more pronounced deviation from ideal linear scaling. The actual computational time for $n_{\rm threads}=64$ is presented in Table~\ref{tab:threads_computational_time_sparce_matrix} for the different system sizes and models.

\subsection{Comparison with Other ED Solvers}
\label{sec:comparison_other_libraries}

Finally, we assess XDiag's performance relative to existing ED libraries. We compare the serial performance of the Lanczos routine against the one in the ALPS library~\cite{Albuquerque2007} and benchmark the sparse matrix generation against the QuSpin library~\cite{Weinberg2017,Weinberg2019}. For both benchmarks, we restrict the comparison to serial performance, as these operations are not parallelized in ALPS or QuSpin.

\begin{figure}
    \centering
    \includegraphics{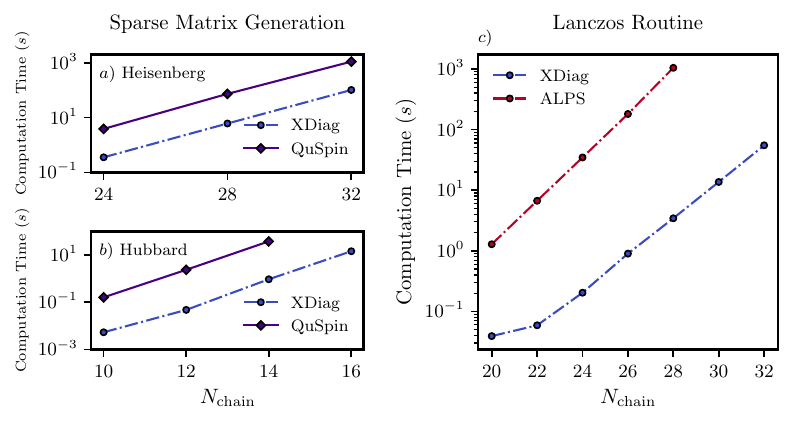}
    \caption{Computational efficiency of XDiag compared to QuSpin and ALPS. Panels $a)$ and $b)$ illustrate the scaling of computational time for generating the Hamiltonian matrix (CSR format) for different system sizes, contrasting XDiag with the QuSpin library. Panel $c)$ benchmarks the Lanczos iteration performance of XDiag against the ALPS library, using a Heisenberg chain in the zero magnetization and momentum sectors across a range of system sizes.}
    \label{fig:comparison_libraries}
\end{figure}

Panels $a)$ and $b)$ of Fig.~\ref{fig:comparison_libraries} illustrate the performance of QuSpin and XDiag in generating the Hamiltonian matrix in CSR format. Panel $a)$ considers the Hamiltonian matrix in the zero magnetization, zero momentum, and even inversion symmetry sectors, while panel $b)$ corresponds to the half-filling, zero momentum, and even inversion symmetry sectors. We observe that the XDiag implementation consistently outperforms QuSpin, with computation times lower by an order of magnitude. Furthermore, XDiag allows for multithreaded parallelization, which can reduce the computational time even further.

In panel $c)$, we compare XDiag's Lanczos performance against that of the ALPS library. The calculations represent the computational time required for five Lanczos iterations on a Heisenberg chain in the zero magnetization and zero momentum sectors (point-group symmetries were not used as they are not currently supported by ALPS). We observe that XDiag performs better than ALPS across all system sizes considered.

\section{Outlook}
\label{sec:outlook}

With XDiag, we strive to create a wide collection of optimized algorithms that enable highly efficient exact diagonalizations of specific quantum many-body systems. In our initial release, we focus on spin $S=1/2$ systems as well as fermionic systems, such as Hubbard and $t$-$J$ models. To efficiently simulate these systems, many algorithmic ideas have been implemented, tested, and benchmarked. Going forward, we want to adhere to a philosophy where we favor efficient implementations over generic implementations. Here, we would like to discuss possible extensions we envision for the future. 

At present, we provide permutational symmetries only with multithreaded parallelization. Several works have previously combined distributed parallelization with permutational symmetries. In Ref.~\cite{Wietek2018}, one of the authors proposed combining sublattice-coding techniques with randomized state distribution to achieve exact diagonalizations up to $N=50$ spin $S=1/2$ systems. Currently, the algorithm rests on so-called sublattice stability, which is satisfied for many one-, two-, and three-dimensional lattices but is not generically available. As such, an implementation for generic interaction graphs would be highly interesting and would facilitate providing these capabilities in a user-friendly way. Interesting approaches have been discussed~\cite{Weisse2013,Westerhout2023} and the algorithms presented there could be a base of an implementation also within XDiag. Combining permutational symmetries with distributed parallelization for electronic models like the $t$-$J$ or Hubbard models, likely, does not require the development of new algorithms as the estimated lookup table sizes for navigating Hilbert spaces are reasonable in this case, as opposed to the spin $S=1/2$ case.

Besides the already provided Hilbert spaces, we plan on adding functionality for other Hilbert spaces as well. An implementation of spinless fermions would be very similar to the implementations of the spin $S=1/2$ system and could adopt the already implemented algorithms to account for additional Fermi signs without major obstacles. Higher spin-$S$ systems require different algorithms to be implemented efficiently. Highly interesting methods have recently been proposed in the context of the DanceQ library~\cite{Schaefer2025a,Schaefer2025b}. There, the authors propose efficient lookup tables, similar to the early Lin tables~\cite{Lin1990} for higher spin $S$ systems and demonstrate impressive performance where an open source implementation is available. As such, these proposals could also be used in a future implementation within the XDiag framework. 

Since we provide routines like \lstinline{apply} to perform on-the-fly matrix-vector multiplication, XDiag could already be used in conjunction with iterative linear algebra libraries such as ARPACK~\cite{arpack}. However, we would still like to provide further iterative algorithms as built-in options within XDiag. Specifically, the locally optimal block preconditioned conjugate gradient (LOBPCG)~\cite{Knyazev2001} is a highly efficient eigensolver with several advantages over the Lanczos algorithm, including the ability to correctly resolve degeneracies. Moreover, linear equation solvers such as the MINRES~\cite{Paige1975} method could also be useful to a wide range of applications. Having built-in implementations of such algorithms would alleviate the difficulties of wrapping them into other third-party iterative libraries. 

\section{Conclusion}
\label{sec:conclusion}
In summary, XDiag represents a state-of-the-art solution for exact diagonalization in quantum many-body physics. Its advanced algorithms, parallelization capabilities, and user-friendly interface make it a valuable tool for researchers seeking to explore the complex behaviors of quantum systems with high precision and efficiency. The full documentation for XDiag is available online~\cite{documentation}, providing comprehensive details on the library's features, installation instructions, and usage examples. XDiag combines the efficiency of advanced algorithms for working in symmetry-adapted bases with the user-friendliness offered by higher-level programming languages. The core library, implemented in C++ for efficiency and expressiveness, is complemented by a comprehensive Julia wrapper, allowing for seamless integration of simulations and data analysis, including visualizations. This dual-language approach ensures that users can leverage the strengths of both C++ and Julia, depending on their specific needs. The library's strong performance is evident in its benchmarks, which demonstrate almost linear scaling in both shared-memory and distributed-memory parallelization scenarios. This efficiency is crucial for handling the exponential growth of the Hilbert space with system size, a common challenge in quantum many-body physics. Looking ahead, XDiag is poised for further enhancements, including the integration of permutational symmetries with distributed parallelization, the addition of new Hilbert spaces, and the implementation of more iterative algorithms. These future developments will continue to advance the field by providing researchers with even more powerful and versatile tools for exact diagonalization.

\section*{Acknowledgements}
We are deeply grateful to Andreas Läuchli for sharing his numerous insights on advanced ED techniques. We thank Andreas Honecker and Sylvain Capponi for numerous discussions and further insights, and Miles Stoudenmire, Matthew Fishman, and Johannes Hofmann for enlightening discussions. Much of the early work on this software was undertaken at the Flatiron Institute, Center for Computational Quantum Physics, a division of the Simons Foundation. 


\paragraph{Funding information}
A.W. acknowledges support by the DFG through the Emmy Noether program (Grant No. 509755282).





\bibliography{main.bib}


\appendix
\section{Operator types}
\label{app:operator_types}
Generic operators in XDiag are represented as \lstinline{OpSum} objects made up of a coupling, which can be a real/complex number or a string, and \lstinline{Op} objects. Every \lstinline{Op} is defined by a `type`. Here we list all the available types implemented in XDiag, their required number of sites, and the blocks for which they are available.

\subsection{List of defined operator types}
\begin{tabularx}{\textwidth}{|>{\ttfamily}l|X|c|>{\ttfamily}X|}
\hline
\textbf{\textrm{Type}} & \textbf{Description} & \textbf{Sites} & \textbf{\textrm{Blocks}} \\
\hline
Hop & A hopping term for $\uparrow$ and $\downarrow$ spins of the form $\textcolor{red}{-}\sum_{\sigma=\uparrow\downarrow} (tc^\dagger_{i\sigma}c_{j\sigma} + \textrm{h.c.})$ & 2 & tJ, Electron, tJDistributed, ElectronDistributed \\
\hline
Hopup & A hopping term for $\uparrow$ spins of the form $\textcolor{red}{-}(tc^\dagger_{i\uparrow}c_{j\uparrow} + \textrm{h.c.})$ & 2 & tJ, Electron, tJDistributed, ElectronDistributed \\
\hline
Hopdn & A hopping term for $\downarrow$ spins of the form $\textcolor{red}{-}(tc^\dagger_{i\downarrow}c_{j\downarrow} + \textrm{h.c.})$ & 2 & tJ, Electron, tJDistributed, ElectronDistributed \\
\hline
HubbardU & A uniform Hubbard interaction across the full lattice of the form $\sum_i n_{i\uparrow}n_{i\downarrow}$ & 0 & Electron, ElectronDistributed \\
\hline
Cdagup & A fermionic creation operator for an $\uparrow$ spin $c^\dagger_{i\uparrow}$ & 1 & tJ, Electron, tJDistributed, ElectronDistributed \\
\hline
Cdagdn & A fermionic creation operator for an $\downarrow$ spin $c^\dagger_{i\downarrow}$ & 1 & tJ, Electron, tJDistributed, ElectronDistributed \\
\hline
Cup & A fermionic annihilation operator for an $\uparrow$ spin $c_{i\uparrow}$ & 1 & tJ, Electron, tJDistributed, ElectronDistributed \\
\hline
Cdn & A fermionic annihilation operator for an $\downarrow$ spin $c_{i\downarrow}$ & 1 & tJ, Electron, tJDistributed, ElectronDistributed \\
\hline
Nup & A number operator for an $\uparrow$ spin $n_{i\uparrow}$ & 1 & tJ, Electron, tJDistributed, ElectronDistributed \\
\hline
Ndn & A number operator for an $\downarrow$ spin $n_{i\downarrow}$ & 1 & tJ, Electron, tJDistributed, ElectronDistributed \\
\hline
Ntot & A number operator $n_i = n_{i\uparrow} + n_{i\downarrow}$ & 1 & tJ, Electron, tJDistributed, ElectronDistributed \\
\hline
Nupdn & double occupancy $d_i = n_{i\uparrow} n_{i\downarrow}$ & 1 & Electron, ElectronDistributed \\
\hline
NupdnNupdn & double occupancy correlation $d_id_j$ & 2 & Electron, ElectronDistributed \\
\hline
NtotNtot & A density-density interaction $n_i n_j$ & 2 & tJ, Electron, tJDistributed, ElectronDistributed \\
\hline
SdotS & A Heisenberg interaction of the form $\mathbf{S}_i \cdot \mathbf{S}_j = S^x_iS^x_j + S^y_iS^y_j + S^z_iS^z_j$ & 2 & Spinhalf, tJ, Electron, SpinhalfDistributed, tJDistributed, ElectronDistributed \\
\hline
SzSz & An Ising interaction of the form $S^z_i S^z_j$ & 2 & Spinhalf, tJ, Electron, SpinhalfDistributed, tJDistributed, ElectronDistributed \\
\hline
Exchange & A spin exchange interaction of the form $\frac{1}{2}(JS^+_i S^-_j + J^*S^-_iS^+_j)$ & 2 & Spinhalf, tJ, Electron, SpinhalfDistributed, tJDistributed, ElectronDistributed \\
\hline
Sz & A local magnetic moment in the $z$-direction $S^z_i$ & 1 & Spinhalf, tJ, Electron, SpinhalfDistributed, tJDistributed, ElectronDistributed \\
\hline
S+ & A local spin raising operator $S^+_i$ & 1 & Spinhalf, SpinhalfDistributed \\
\hline
S- & A local spin lowering operator $S^-_i$ & 1 & Spinhalf, SpinhalfDistributed \\
\hline
ScalarChirality & A scalar chirality interaction of the form $\mathbf{S}_i \cdot (\mathbf{S}_j \times \mathbf{S}_k)$ & 3 & Spinhalf \\
\hline
tJSzSz & An Ising interaction as encountered in the $t-J$ model of the form $S^z_i S^z_j - \frac{n_i n_j}{4}$ & 2 & tJ, tJDistributed \\
\hline
tJSdotS & An Heisenberg interaction as encountered in the $t-J$ model of the form $\mathbf{S}_i \cdot \mathbf{S}_j - \frac{n_i n_j}{4}$ & 2 & tJ, tJDistributed \\
\hline
Matrix & A generic spin interaction no an arbitrary number of sites defined via a coupling matrix & arbitrary & Spinhalf \\
\hline
\end{tabularx}

\end{document}